\documentclass{emulateapj}
\usepackage{natbib}
\usepackage{psfig}
\usepackage{epsfig}

\begin{document}
\title{Numerical $3+1$ General Relativistic Magnetohydrodynamics:
A local characteristic approach } 

\author{Luis Ant\'on\altaffilmark{1}, 
Olindo Zanotti\altaffilmark{1}, 
Juan A. Miralles\altaffilmark{2},
Jos\'e M$^{\underline{\mbox{a}}}$ Mart\'{\i}\altaffilmark{1},
\\ 
Jos\'e M$^{\underline{\mbox{a}}}$ Ib\'a\~nez\altaffilmark{1}, 
Jos\'e A. Font\altaffilmark{1},
Jos\'e A. Pons\altaffilmark{2}}

\altaffiltext{1}{Departamento de Astronom\'{\i}a y Astrof\'{\i}sica,
Universidad de Valencia, Edificio de Investigaci\'on, Dr. Moliner 50,
46100 Burjassot (Valencia), Spain}
\altaffiltext{2}{Departament de F\'{\i}sica Aplicada,
Universitat d'Alacant, Ap. Correus 99,
03080 Alacant, Spain}

%\date{Received.....; accepted.....} 

\markboth{Ant\'on et al.: General relativistic magnetohydrodynamics}{} 

\begin{abstract}

We present a general procedure to solve numerically the general relativistic 
magnetohydrodynamics (GRMHD) equations within the framework of the $3+1$ 
formalism. The work reported here extends our previous investigation in
general relativistic hydrodynamics \citep{banyuls:97} where magnetic 
fields were not considered. The GRMHD equations are written in conservative 
form to exploit their hyperbolic character in the solution procedure. All 
theoretical ingredients necessary to build up high-resolution shock-capturing 
schemes based on the solution of local Riemann problems (i.e.~Godunov-type 
schemes) are described. In particular, we use a
renormalized set of regular eigenvectors of the flux
Jacobians  of the relativistic magnetohydrodynamics equations.
In addition, the paper describes a procedure 
based on the equivalence principle of general relativity that allows the use 
of Riemann solvers designed for special relativistic magnetohydrodynamics
in GRMHD. 
Our formulation and numerical methodology are assessed by
performing various test  simulations 
recently considered by different authors. 
These include  magnetized
shock tubes, spherical accretion onto a  Schwarzschild
black hole, equatorial accretion onto a Kerr black
hole, and  magnetized thick accretion disks around a
black hole prone to the magnetorotational instability.

\end{abstract}


%%%%%%%%%%%%%%%%%%%%%%%%%%%%%%%%%%%%%%%%%%%%%%%%%%
\section{Introduction}
\label{intro}
%%%%%%%%%%%%%%%%%%%%%%%%%%%%%%%%%%%%%%%%%%%%%%%%%%

In several astrophysical scenarios both magnetic and gravitational
fields play an important role in determining the evolution of the
matter. In these scenarios it is a common fact the presence of compact
objects such as neutron stars, most of which have intense magnetic
fields of order $10^{12}-10^{13}$G, or even larger at birth, $\sim
10^{14}-10^{15}$G, as inferred from studies of anomalous X-ray pulsars
and soft gamma-ray repeaters~\citep{kouveliotou}. In some cases, i.e. 
in the so-called magnetars, the magnetic fields can be so strong to 
affect the internal structure of the star \citep{bocquet}. In a different 
context, the most promising mechanisms for producing relativistic jets 
like those observed in AGNs and microquasars, and the ones conjectured 
to explain gamma-ray bursts involve the hydromagnetic centrifugal 
acceleration of material from an accretion disk, or the extraction of
rotational energy from the ergosphere of a Kerr black 
hole~\citep{penrose,blandford:77,blandford:82}.  In addition, the 
differential rotation of the magnetized plasma in the disk is responsible 
of the magnetorotational instability, which plays an important role in 
transporting angular momentum outward~\citep{balbus}.  

If the gravitational field is strong enough, as in the vicinity of a
compact object, the Newtonian description of gravity is only a rough
approximation and general relativity becomes necessary. In such a theory,
the so called 3+1 formalism ~\citep{ADM} has proved particularly 
useful for numerical simulations involving time-dependent computations of
hydrodynamical flows in curved spacetimes, either static or dynamic. The 
interested reader is addressed to \citet{fontlr} and references therein 
for an up-to-date overview of the different approaches that have been 
introduced during the years for solving the general relativistic hydrodynamics 
equations.

% Numerical GR-MHD

On the other hand, the inclusion of magnetic fields and the development 
of mathematical formulations of the magnetohydrodynamic (MHD) equations 
in a form suitable for efficient numerical implementations is still in 
an exploratory phase, although considerable progress has already been 
achieved in the last few years.

Numerical studies in special relativistic
magnetohydrodynamics (SRMHD)
have been undertaken by a growing 
number of authors~\citep{komissarov99,balsara01,koldoba,delzanna,tobias}. In 
particular, \citet{komissarov99}, \citet{balsara01}, and \citet{koldoba} 
developed independent {\it upwind} high-resolution shock-capturing (HRSC)
schemes (also referred to as Godunov-type schemes), providing the 
characteristic information of the corresponding system of equations, which 
is the crucial building block in such type of schemes. In addition, 
\citet{komissarov99} and \citet{balsara01} proposed a comprehensive sample 
of tests to validate numerical MHD codes in special relativity (SR). Recently, 
\citet{delzanna} have developed a third order shock-capturing {\it central} 
scheme for SRMHD which sidesteps the use of Riemann solvers in the solution 
procedure (see, e.g.~\citet{toro:97} for general definitions on HRSC schemes).
Simulations of the morphology and dynamics of magnetized relativistic jets with
Godunov-type schemes have been reported by \citet{tobias}. In addition, the
exact solution of the Riemann problem in SRMHD, for some
particular orientation of the magnetic field and the fluid velocity field,
has been obtained by \citet{romero}.

Correspondingly, 3+1 representations of GRMHD were first analyzed 
by \citet{sloan85}, \citet{evans88}, \citet{zhang}, \citet{yokosawa93}, and, 
more recently, by \citet{koide98}, \citet{devilliers1}, \citet{baumgarte1},
\citet{gammie:03}, \citet{komissarov05},  \citet{duez05}
and \citet{shibata:05}. Most 
of the existing applications to date are in the field of black hole accretion 
and jet formation. In~\cite{yokosawa93,yokosawa95} the transport of energy and
angular momentum in magneto-hydrodynamical accretion onto a rotating black 
hole was studied adopting Wilson's formulation for the hydrodynamic equations 
\citep{wilson79}, conveniently modified to account for the magnetic terms. The 
magnetic induction equation was solved using the constrained transport method 
of~\cite{evans88}. Later on, Koide and coworkers performed the first MHD 
simulations of jet formation in general relativity \citep{koide98,koide00} 
in the context of the Blandford-Payne mechanism. These authors solved the
MHD equations in the test-fluid approximation (in the background geometry of 
Schwarzschild/Kerr spacetimes) using a second-order finite difference central scheme with 
nonlinear dissipation. Employing the same numerical approach \citet{koide02a} 
and \citet{koide03} studied the validity of the so-called MHD Penrose 
process to extract rotational energy from a Kerr black hole by simulating 
the evolution of a rarefied plasma with a uniform magnetic field. 
\citet{komissarov05} has also recently investigated this topic finding
evidence in favour of the extraction of rotational energy of the black
hole by the Blandford-Znajek mechanism \citep{blandford:77} but against 
the development of strong relativistic outflows or jets. The long term 
solution found by \citet{komissarov05} shows properties which are significantly 
different from those of the short initial (transient) phase studied by 
\citet{koide03}. An additional astrophysical application in the context of 
electromagnetic extraction of energy from a Kerr black hole is represented 
by the analysis of \cite{mckinney:04}, who have compared the analytic prediction 
of \cite{blandford:77} with time evolution calculations. Finally, two
different groups \citep{devilliers1,devilliers:03,gammie:03} have started 
programs to investigate the time-varying behaviour of magnetized accretion 
flows onto Kerr black holes, with great emphasis on the issue of the development 
of the magnetorotational instability in thick accretion disks (see also
\citet{yokosawa05}). While \cite{devilliers1,devilliers:03} adopt a 
nonconservative (ZEUS-like) scheme, the approach followed by \cite{gammie:03} 
is based on a conservative HRSC scheme, namely the so-called HLL scheme 
of \citet{harten:83}. 

% Our goal

To the light of the existing literature on the subject it is clear that 
astrophysical applications of Godunov-type schemes in general relativistic MHD 
have only very recently been reported \citep{gammie:03,komissarov05,duez05}.  
Our goal in this paper is to present the evolution equations for the magnetic 
field and for the fluid within the 3+1 formalism, formulated in a suitable way 
to apply Godunov-type schemes based on (approximate) Riemann solvers. Our 
numerical procedure uses two original ingredients. On the one hand, the code 
incorporates a local coordinate transformation to Minkowskian coordinates, 
similar to the one developed for relativistic hydrodynamics in \citet{pons:98},
prior to the computation of the numerical fluxes. In this way, Riemann solvers 
designed for SRMHD can be straightforwardly used in GRMHD calculations. We 
note that \citet{komissarov05} applies the same approach, using a HRSC scheme 
based on the SR Riemann solver described in \citet{komissarov99} and adapted 
to general relativity following the procedure laid out in \citet{pons:98}.
We present here, however, a number of tests assessing the feasibility of the 
approach. As a second novel ingredient, we use a {\it renormalized} set of 
right and left eigenvectors of the flux vector Jacobians of the SRMHD equations 
which are regular and span a complete basis in any physical state, including 
degenerate states. 

The organization of the paper is as follows. We start by introducing
the mathematical framework in \S 2, including the essentials of the
3+1 formalism, the description of the magnetic field, the induction
equation and the conservation equations of particle number, and
stress-energy tensor in conservative form. A brief analysis of the
hyperbolic structure of the GRMHD system of equations is given in
\S3. The numerical procedure to solve the equations is described in \S
4. Finally, in \S 5 we present the results of some numerical tests and
applications in order to assess our formulation and methodology. The
summary of our work is given in \S 6. Throughout the paper Latin indices 
run from 1 to 3 and Greek indices from 0 to 3. Four-vectors are indistinctly 
denoted using index notation or boldface letters, e.g. $u^{\mu}$, ${\bf u}$.  
We adopt geometrized units by setting $c=G=1$.

%%%%%%%%%%%%%%%%%%%%%%%%%%%%%%%%
\section{Mathematical framework}
\label{I}
%%%%%%%%%%%%%%%%%%%%%%%%%%%%%%%%

%%%%%%%%%%%%%%%%%%%%%%%%%%%%%%%%%%%%%%%%%%%%%%%%%%%%%%%
\subsection{The Eulerian observer in the 3+1 formalism}
%%%%%%%%%%%%%%%%%%%%%%%%%%%%%%%%%%%%%%%%%%%%%%%%%%%%%%%

In the 3+1 formalism the line element of the spacetime can be
written as
\begin{equation}
ds^{2} = -(\alpha^{2}-\beta_{i}\beta^{i}) dt^{2}+
2 \beta_{i} dx^{i} dt + \gamma_{ij} dx^{i}dx^{j}, 
\end{equation}
where $\alpha$ (lapse function), $\beta^i$ (shift vector) and
$\gamma_{ij}$ (spatial metric) are functions of the coordinates $t$,
$x^i$. A natural observer associated with the 3+1 splitting is the one 
with four velocity ${\bf n}$ perpendicular to the hypersurfaces of 
constant $t$ at each event in the spacetime. This is the so-called 
{\em Eulerian observer}\footnote{In  the Kerr metric this Eulerian observer is
indeed the observer with zero azimuthal angular momentum (ZAMO) as measured
from infinity.}.  
The contravariant and covariant components 
of ${\bf n}$ are given by
\begin{equation}
n^\mu=\frac{1}{\alpha}(1,-\beta^i),
\end{equation}
and
\begin{equation}
n_\mu=(-\alpha,0,0,0) , \
\end{equation}
respectively. 
In spacetimes containing matter an additional natural 
observer is the one that follows the fluid during its motion, also called 
the {\em comoving observer}, with four-velocity ${\bf u}$.  With the standard
definition, the three-velocity of the fluid as measured by the
Eulerian observer can be expressed as
\begin{equation}
\label{3vel}
v^i\equiv \frac{h^i_\mu u^\mu}{-{\bf u}\cdot {\bf n}},
\end{equation}
where $-{\bf u}\cdot {\bf n}\equiv W$ is the relative Lorentz factor
between ${\bf u}$ and ${\bf n}$, while $h_{\mu\nu}=g_{\mu\nu} +
n_\mu n_\nu$ is the the projector onto the hypersuface orthogonal
to ${\bf n}$, whose spatial terms are given by $h_{ij}=\gamma_{ij}$.  
From Eq.~(\ref{3vel}) it follows that
\begin{equation}
v^i=\frac{u^i}{\alpha u^t}+\frac{\beta^i}{\alpha} \ , 
\end{equation}
while $v_i=u_i/W$. Note that the Lorentz factor satisfies the relation 
$W=1/\sqrt{(1-v^2)}=\alpha u^t$, where $v^2=\gamma_{ij}v^iv^j$ is the 
squared modulus of the three-velocity of the fluid with respect to the 
Eulerian observer.

%%%%%%%%%%%%%%%%%%%%%%%%%%%%%%%%%%%%%%%%%%%%%%%%%%
\subsection{Magnetic field evolution}
%%%%%%%%%%%%%%%%%%%%%%%%%%%%%%%%%%%%%%%%%%%%%%%%%%

A complete description of the electromagnetic field in general
relativity is provided by the Faraday electromagnetic tensor field
$F^{\mu\nu}$.  This tensor is related to the electric and magnetic
field, $E^\mu$ and $B^\mu$, measured by a generic observer with
four-velocity $U^\mu$, as follows,
\begin{equation}
F^{\mu\nu}=U^\mu E^\nu- U^\nu E^\mu -
\eta^{\mu\nu\lambda\delta} U_\lambda B_\delta,
\label{eq:faraday}
\end{equation}
$\eta^{\mu\nu\lambda\delta}$ being the volume element, 
\begin{equation}
\eta^{\mu\nu\lambda\delta}=\frac{1}{\sqrt{-g}}
[\mu\nu\lambda\delta],
\end{equation}
where $g$ is the determinant of the 4-metric ($g=\det{g_{\mu\nu}}$)
and $[\mu\nu\lambda\delta]$ is the completely antisymmetric
Levi-Civita symbol.  Both, ${\bf E}$ and ${\bf B}$ are orthogonal to
${\bf U}$, ${\bf E}\cdot {\bf U}={\bf B}\cdot{\bf U}=0$.  The dual of
the electromagnetic tensor $^*F^{\mu\nu}$ is defined as
\begin{equation}
^*F^{\mu\nu}=\frac{1}{2}\eta^{\mu\nu\lambda\delta}
F_{\lambda\delta},
\end{equation}
and in terms of the electric and  magnetic field measured by the observer 
${\bf U}$ is given by
\begin{equation}
^*F^{\mu\nu}=U^\mu B^\nu- U^\nu B^\mu +
\eta^{\mu\nu\lambda\delta} U_\lambda E_\delta.
\end{equation}
From these equations, ${\bf E}$ and ${\bf B}$ can be expressed in
terms of the electromagnetic tensor and the four-velocity ${\bf U}$ as
follows
\begin{eqnarray}
\label{emu}
E^\mu&=&F^{\mu\nu}U_\nu, \\ 
\label{bmu}
B^\mu&=&^*F^{\mu\nu}U_\nu.
\end{eqnarray}
In terms of the electromagnetic tensor, Maxwell's equations are
written as follows,
\begin{eqnarray}
\label{max_1}
\nabla_\nu\, ^*F^{\mu\nu}&=&0,  \\
\label{max_2}
\nabla_\nu F^{\mu\nu}&=&4\pi{\cal J}^\mu,
\end{eqnarray}
where $\nabla_\nu$ stands for the covariant derivative and ${\cal
J}^\mu$ is the electric four-current. According to Ohm's
law, the latter can be in general expressed as
\begin{equation}
{\cal J}^\mu=\rho_q u^\mu + \sigma F^{\mu\nu}u_\nu,
\end{equation}
where $\rho_q$ is the proper charge density measured by the comoving
observer and $\sigma$ is the
electric conductivity.  Maxwell's equations can be further simplified
if one assumes that the fluid is a perfect conductor. In this case the
fluid has infinite conductivity and, in order to keep the current
finite, the term proportional to the conduction current,
$F^{\mu\nu}u_\nu$, must vanish, which means that the electric field
measured by the comoving observer is zero.  This case corresponds to
the so-called ideal MHD condition.  We can take
advantage of this condition to express the electric field measured by
the observer ${\bf U}$ as a function of the magnetic field ${\bf B}$
measured by the same observer and of the four-velocities $U^\mu$ and
$u^\mu$. Straightforward calculations give
\begin{equation}
E^\mu=\frac{1}{W}\eta^{\mu\nu\lambda\delta}u_\nu U_\lambda B_\delta.
\label{EbyU}
\end{equation}
If we choose ${\bf U}$ as the four-velocity of the Eulerian observer, 
${\bf U}={\bf n}$, Eq.~(\ref{EbyU}) provides
\begin{eqnarray}
E^0&=&0, \\
E^i&=&-\alpha \eta^{0ijk} v_j B_k,
\end{eqnarray}
or, in terms of three-vectors, $\vec{E}=- \vec{v}\times\vec{B}$, where
the arrow means that the vector lies in the `absolute space' and the
cross product is defined using the induced volume element in the
absolute space $\eta^{ijk}= \alpha \eta^{0ijk}$.  Using the above
relations, the dual of the electromagnetic field can be written 
in terms of the magnetic field only
\begin{equation}
^*F^{\mu\nu}=\frac{u^\mu B^\nu-u^\nu B^\mu}{W},
\end{equation}
and Maxwell's equations $\nabla_\nu ^*F^{\mu\nu}=0$ reduce to the 
divergence-free condition plus the induction equation for the
evolution of the magnetic field
\begin{eqnarray}
\frac{\partial (\sqrt{\gamma} B^i)}{\partial x^i} &=&0, 
\label{divfree} \\
\frac{1}{\sqrt{\gamma}}\frac{\partial}{\partial t} (\sqrt{\gamma}
B^i)&=&\frac{1}{\sqrt{\gamma}} \frac{\partial}{\partial
x^j}\{\sqrt{\gamma} [(\alpha v^i-\beta^i)B^j \nonumber \\ 
&&-(\alpha
v^j-\beta^j)B^i]\},
\label{eq:evB}
\end{eqnarray}
or, in terms of three-vectors,
\begin{eqnarray}
\vec{\nabla}\cdot\vec{B}&=&0 \\ 
\frac{1}{\sqrt{\gamma}}\frac{\partial
}{\partial t} \left(\sqrt{\gamma}\vec{B}\right)& =
&\vec{\nabla}\times\left[\left(\alpha\vec{v}-
\vec{\beta}\right)\times\vec{B}\right].
\label{evolB}
\end{eqnarray}

%%%%%%%%%%%%%%%%%%%%%%%%%%%%%%%%%%%%%%%%%%%%%%%%%%
\subsection{Conservation Equations}
\label{conservation}
%%%%%%%%%%%%%%%%%%%%%%%%%%%%%%%%%%%%%%%%%%%%%%%%%%

Once we have established the magnetic field evolution equation in the 
ideal MHD case, we need to obtain the evolution equations for the matter 
fields. These equations can be expressed as the local conservation laws 
of baryon number and energy-momentum. For the baryon number we have
\begin{equation}
\nabla_\nu J^\nu=0,
\label{eq:evrho}
\end{equation}
${\bf J}$ being the rest-mass current, $J^\mu=\rho u^\mu$, where $\rho$
denotes the rest-mass density. The conservation of the
energy-momentum is given by
\begin{equation}
\nabla_\nu T^{\mu\nu}=0,
\label{eq:evtmunu}
\end{equation}
where $T^{\mu\nu}$ is the energy-momentum tensor. For a fluid endowed with a 
magnetic field, this tensor is obtained by adding the energy-momentum tensor 
of the fluid to that of the electromagnetic field:
\begin{equation}
T^{\mu\nu}=T_{\rm Fluid}^{\mu\nu}+T_{\rm EM}^{\mu\nu} \ .
\end{equation}
When the fluid is assumed to be perfect, $T_{\rm Fluid}^{\mu\nu}$ is given 
by
\begin{equation}
T_{\rm Fluid}^{\mu\nu}=\rho h u^\mu u^\nu + p g^{\mu\nu},
\end{equation}
where $g_{\mu\nu}$ is the metric, $p$ is the pressure, and $h$ is the specific 
enthalpy, defined by $h=1+\varepsilon +p/\rho$, $\varepsilon$ being the 
specific internal energy. The fluid is further assumed to be in local 
thermodynamic equilibrium, and there exists an equation of state of the form 
$p=p(\rho,\varepsilon)$ which relates the pressure with $\rho$ and $\varepsilon$. 
On the other hand, the energy-momentum tensor $T_{\rm EM}^{\mu\nu}$ of the 
electromagnetic field can be obtained from the
electromagnetic tensor, ${\bf F}$,  as follows  
\begin{equation}
\label{T_em1}
T_{\rm EM}^{\mu\nu}=\frac{1}{4\pi}\left(F^{\mu\lambda}F^\nu_{\hspace{0.2cm}\lambda} -
\frac{1}{4}g^{\mu\nu} F^{\lambda\delta}F_{\lambda\delta}\right).
\end{equation}
Furthermore, from Eq.~(\ref{eq:faraday}) and exploiting the ideal
MHD condition, 
the electromagnetic tensor can be expressed in terms of the magnetic
field $b^\mu$ measured by the comoving observer as
\begin{equation}
F^{\mu\nu} = -\eta^{\mu\nu\lambda\delta}u_\lambda b_\delta,
\end{equation}
and Eq.~(\ref{T_em1}) can be rewritten as
\begin{equation}
T_{\rm EM}^{\mu\nu}=\left(u^\mu u^\nu+\frac{1}{2}
g^{\mu\nu}\right)b^2 - b^\mu b^\nu,
\end{equation}
where $b^2=b^\nu b_\nu$ and where the
magnetic field four vector has been redefined by dividing it 
by the factor $\sqrt{4\pi}$.
As a result, the total energy-momentum tensor,
fluid plus electromagnetic field, is given by
\begin{equation}
T^{\mu\nu}=\rho h^* u^\mu
u^\nu+p^* g^{\mu\nu}- b^\mu b^\nu.
\end{equation}
where we have introduced the definitions $p^*=p+b^2/2$ and $h^*=h+b^2/\rho$. 
Note that if we consistently define $\varepsilon^*=\varepsilon+b^2/(2\rho)$, 
the following relation, $h^*=1+\varepsilon^*+p^*/\rho$, is fulfilled. 

In order to write the evolution equations (\ref{eq:evrho}), (\ref{eq:evtmunu}) 
in a conservation form suitable for numerical applications, let us define a 
basis adapted to the Eulerian observer,
\begin{equation}
{\bf e}_{(\lambda)}=\{{\bf n},\partial_i\},
\end{equation}
where $\partial_i$ are the coordinate vectors that are tangent to the 
hypersurface $t$=const, and, therefore, ${\bf n}\cdot
\partial_i=0$. 
This allows us to define the following five 4-vectors
${\cal D}_{(A)}$: 
\begin{equation}
{\cal D}_{(A)}=\{{\bf T}({\bf e}_{(\lambda)}, \cdot),{\bf J}\},\hspace {1 cm} 
A=0,\dots,4.
\end{equation}
Hence the above system of equations (\ref{eq:evrho}),
(\ref{eq:evtmunu}) can be written as
\begin{equation}
\nabla_\nu{\cal D}_{(A)}^\nu=s_{(A)} , \
\end{equation}
where the five quantities $s_{(A)}$ on the right-hand side
--{\it the sources}--, are
\begin{equation}
s_{(A)}=\{T^{\alpha\beta}\nabla_\mu e_{(\lambda)\nu},0\}
\ .
\end{equation}
The covariant derivatives of the basis vectors,
$\nabla_\mu e_{(\lambda)\nu}$, are obtained in the usual
manner as
\begin{equation}
\nabla_\mu e_{(\lambda)\nu}=\frac{\partial e_{(\lambda)\nu}}{\partial x^\mu}-
\Gamma^\delta_{\nu\mu}e_{(\lambda)\delta},
\end{equation}
where $\Gamma^\delta_{\nu\mu}$ are the Christoffel symbols, and
\begin{equation}
e_{(0)\nu}=-\alpha \delta_{0\nu}, \hspace{0.5 cm} e_{(k)\nu}=g_{k\nu}=
(\beta_k,\gamma_{kj}).
\end{equation}
In a similar way to the pure hydrodynamics case \citep{banyuls:97},
if we now define the following quantities measured by an Eulerian observer,
\begin{eqnarray}
\label{conv_1}
D&\equiv& - J_\nu n^\nu=\rho W \\
\label{conv_2}
S_j &\equiv& - {\bf T}({\bf n},{\bf
e}_{(j)})=\rho h^* W^2 v_j - \alpha b^0 b_j  \\
\label{conv_3}
\tau&\equiv& {\bf T}({\bf n}, {\bf n})=\rho h^* W^2-p^* - \alpha^2(b^0)^2 - D
\end{eqnarray}
i.e.~the rest-mass density, the momentum density of the magnetized
fluid in the $j$-direction, and its total energy density (subtracting
the rest-mass density in order to consistently recover the Newtonian limit), 
respectively, the system of GRMHD equations can be written explicitly in
conservative form. Together with the equation for the evolution of the
magnetic field as measured by the Eulerian observer,
Eq.~(\ref{eq:evB}), the fundamental GRMHD system of equations can be
written in the following general form
\begin{equation}
\frac{1}{\sqrt{-g}} \left(
\frac {\partial \sqrt{\gamma}{\bf F}^{0}}
{\partial x^{0}} +
\frac {\partial \sqrt{-g}{\bf F}^{i}}
{\partial x^{i}} \right)
 = {\bf S},
\label{eq:fundsystem}
\end{equation}
where the quantities ${\bf F}^{\mu}$ (${\bf F}^0$ being the state vector and
${\bf F}^i$ being the fluxes) are  
\begin{eqnarray}
{\bf F}^0 & =& \left[\begin{array}{c}
D \\
S_j \\
\tau \\
B^k
\end{array}\right],
\label{state_vector}
\end{eqnarray}
\begin{eqnarray}
{\mathbf F}^i & =& \left[\begin{array}{c}
D \tilde{v}^i\\
S_j \tilde{v}^i + p^{*} \delta^i_j - b_j B^i/W \\
\tau \tilde{v}^i + p^{*} v^i - \alpha b^0 B^i/W \\
\tilde{v}^i B^k-\tilde{v}^k B^i
\end{array}\right]
\label{flux2}
\end{eqnarray}
%
%\begin{eqnarray}
%{\mathbf F}^i & =& \left[\begin{array}{c}
%D \tilde{v}^i\\
%\rho h^* W^2 v_j \tilde{v}^i + p^*\delta^i_j -b^i b_j\\
%\rho h^*W^2 \tilde{v}^i  +p^*\beta^i/\alpha
%- D\tilde{v}^i - \alpha b^0 b^i \\
%\tilde{v}^i B^k-\tilde{v}^k B^i
%\end{array}\right],
%\end{eqnarray}
%
with $\tilde{v}^i=v^{i}-\frac{\beta^i}{\alpha}$.
The corresponding sources ${\bf S}$ are given by
\begin{eqnarray}
{\mathbf S} & =& \left[\begin{array}{c}
0 \\
T^{\mu \nu} \left(
\frac {\partial g_{\nu j}}{\partial x^{\mu}} -
\Gamma^{\delta}_{\nu \mu} g_{\delta j} \right) \\
\alpha  \left(T^{\mu 0} \frac {\partial {\rm ln} \alpha}{\partial x^{\mu}} -
T^{\mu \nu} \Gamma^0_{\nu \mu} \right) \\
0^k
\end{array}\right],
\end{eqnarray}
where  $0^k \equiv(0,0,0)^T$.  
Note that the following fundamental relations hold
between the four components of the magnetic field in the comoving
frame, $b^\mu$, and the three vector components $B^i$ measured by the
Eulerian observer:
\begin{eqnarray}
\label{b0}
b^0 &=& \frac{WB^iv_i}{\alpha} \\
\label{bi}
b^i &=& \frac{B^i + \alpha b^0 u^i}{W} \ .
\end{eqnarray}
Finally, the modulus of the magnetic field can be written as
\begin{equation}
  b^2 = \frac{B^2 + \alpha^2 (b^0)^2}{W^2} \ ,
\end{equation}
where $B^2 = B^iB_i$. 

%%%%%%%%%%%%%%%%%%%%%%%%%%%%%%%%%%%%%%%%%%%%%%%%%%
\section{Hyperbolic structure}
\label{II}
%%%%%%%%%%%%%%%%%%%%%%%%%%%%%%%%%%%%%%%%%%%%%%%%%%

In Section~\ref{conservation} we have written the GRMHD equations 
in conservative form anticipating the use of numerical methods 
specifically designed to solve conservation equations, as will be 
explained in the next Section. These methods
strongly rely on the hyperbolic character of the equations and on the
associated wave structure. Following \citet{anile}, in order to
analyze the hyperbolicity of the equations it is convenient to
write them in a more suitable form. If we take the following
set of variables, ${\bf V} = (u^\mu,b^\mu,p,s)$, where $s$ is the
specific entropy, the system of equations can be written as a
quasi-linear system of the form
\begin{equation}
{\cal A}^{\mu A}_B\nabla_\mu V^B=0,
\label{amuab}
\end{equation}
where, $A$ and $B$ run from 0 to 9, as the number of variables, and
the $10\times 10$ matrices ${\cal A}^{\mu}$ are given by
\begin{eqnarray}
{\cal A}^{\mu} = \left( \begin{array}{cccc} {\cal C}
u^\mu \delta^{\alpha}_{ \beta}\; & -b^{\mu}\delta^{\alpha}_{\beta} +
P^{\alpha\mu}b_\beta & l^{\alpha\mu} & 0^{\alpha\mu} \\ 
b^\mu \delta^{\alpha}_{ \beta} & -u^\mu\delta^{\alpha}_{\beta} &
f^{\mu\alpha} & 0^{\alpha\mu} \\ 
\rho h \delta^{\mu}_\beta & 0^{\mu}_\beta & u^\mu/c_s^2 & 0^\mu \\
0^{\mu}_\beta & 0^{\mu}_\beta & 0^\mu & u^\mu \\ 
\end{array} \right)
\label{amu}
\end{eqnarray} 
where $c_s$ stands for the speed of sound 
\begin{eqnarray}
c_s^2=\left(\frac{\partial p}{\partial e}\right)_s,
\end{eqnarray}
$e$ being the mass-energy density of the fluid $e=\rho(1+\varepsilon)$.
In Eq.~(\ref{amu}) the following definitions are introduced:
\begin{eqnarray}
{\cal C}&=&\rho h + b^2,
\\
P^{\alpha\mu}&=&g^{\alpha\mu}+2 u^\alpha u^\mu, 
\\
l^{\mu\alpha}&=&(\rho h g^{\mu\alpha}+(\rho h -b^2/c_s^2) u^\mu u^\alpha)/
\rho h, \\ 
f^{\mu\alpha}&=&(u^\alpha b^\mu/c_s^2- u^\mu b^\alpha)/\rho h,
\end{eqnarray}
as well as the notation
\begin{equation}
0^\mu \equiv 0, \,\,\,\, 0^{\alpha \mu} \equiv (0,0,0,0)^{\rm T}, \,\,\,\,
0^\mu_\beta \equiv (0,0,0,0).
\end{equation}
If $\phi(x^\mu)=0$ defines a characteristic hypersurface of the above system 
(\ref{amuab}), the characteristic matrix, given by ${\cal A^\epsilon}\phi_\epsilon$ can
be written as
\begin{eqnarray}
\label{ch_matrix}
{\cal A}^{\epsilon} \phi_{\epsilon} = \left( \begin{array}{cccc}
{\cal C} a \delta^{\mu}_{ \nu}  &  m^{\mu}_{\nu} & l^{\mu} & 0^{\mu} \\
\mathcal{B} \delta^{\mu}_{ \nu}  & -a \delta^{\mu}_{\nu} & f^{\mu} & 0^{\mu} \\
\rho h \phi_{\nu} &  0_{\nu} &  a/c_s^2 & 0 \\
 0_{\nu} &  0_{\nu} & 0 &  a \\ 
\end{array}\right)
\end{eqnarray}
where $ \phi_\mu = \nabla_\mu \phi$, $a = u^{\mu} \phi_{\mu}$,
$\mathcal{B}= b^{\mu} \phi_{\mu}$, 
$l^{\mu}= l^{\mu\nu}
\phi_\nu=\phi^\mu+(\rho h - b^2/c_s^2) a u^\mu /\rho h+
\mathcal{B} b^\mu/\rho h$, $f^{\mu}= f^{\mu\nu}\phi_\nu
=(a b^\mu/c_s^2-\mathcal{B} u^\mu)/\rho h$, and
$m^\mu_\nu=(\phi^\mu+2au^\mu)b_\nu-\mathcal{B}\delta^\mu_\nu$.  
The determinant of the matrix (\ref{ch_matrix}) must
vanish, i.e.
\begin{equation}
{\rm det}({\cal A}^{\mu} \phi_{\mu})={\cal C}\,a^2
\mathcal{A}^2 {\cal N}_4 = 0 \ ,
\label{eq:det}
\end{equation}
where
\begin{eqnarray}
 {\cal A} &=& {\cal C} a^2 -\mathcal{B}^2, \\
\label{N4}
 {\cal N}_4 &=& \rho h \left( \frac{1}{c_s^2} -1 \right) a^4 -
  \left(\rho h +\frac{b^2}{c_s^2} \right) a^2 G
 +\mathcal{B}^2 G \ ,
\end{eqnarray}
and $G = \phi^{\mu}\phi_{\mu}$.
If we now consider a wave propagating in an arbitrary direction $x$ with a
speed $\lambda$, the normal to the characteristic hypersurface is
given by the four-vector
\begin{equation}
\label{ppp}
\phi_\mu=(-\lambda,1,0,0), \
\end{equation}
and by substituting Eq.~(\ref{ppp}) in Eq.~(\ref{eq:det}) we obtain
the so called {\it characteristic polynomial}, whose zeroes give the
characteristic speed of the waves propagating in the $x$-direction.
Three different kinds of waves can be obtained according to which
factor in equation~(\ref{eq:det}) becomes zero. For entropic waves
$a=0$, for Alfv\'en waves $\mathcal{A}=0$, and for magnetosonic waves
${\cal N}_4=0$.  

Let us next analyze in more detail the characteristic equation. First of 
all, since the four-vector $\phi_\mu$ must be spacelike (this is a property of 
the RMHD system of equations~\citep{anile}), it follows that $\phi^\mu\phi_\mu>0$. 
In terms of the wave speed $\lambda$ we obtain
\begin{equation}
-\alpha\sqrt{\gamma^{xx}}-\beta^x< \lambda < 
	  \alpha \sqrt{\gamma^{xx}} - \beta^x.
\end{equation}
The characteristic speed $\lambda$ of the entropic waves propagating
in the $x$-direction, given by the solution of the equation $a=0$, is
the following
\begin{equation}
\lambda=\alpha v^x - \beta^x.
\end{equation}
For Alfv\'en waves, given by $\mathcal{A}=0$, there are two solutions 
corresponding, in general, to different speeds of the waves,
\begin{equation}
\lambda=\frac{b^x\pm\sqrt{{\cal C}}u^x}
{b^0\pm\sqrt{{\cal C}}u^t}.
\end{equation}
% ----------------------------------------------------
% I THINK THAT THE FOLLOWING EQUATION IS THE CORRECT
%  ONE. PLEASE UNCOMMENT IF YOU AGREE. FOR SURE IS THE
%  ONE USED IN THE CODE...
%\begin{equation}
%\lambda=\frac{b^x\pm\sqrt{{\cal C}}u^x } {b^0\pm\sqrt{{\cal C}}W/\alpha} 
% \end{equation}
%----------------------------------------------------
%
In the case of magnetosonic waves it is however not possible, in general, to
obtain explicit expressions for their speeds since they are given by
the solutions of the quartic equation ${\cal N}_4=0$
 with $a$, ${\cal B}$ and $G$ explicitly
written in terms of $\lambda$ as
\begin{eqnarray}
a &=& \frac{W}{\alpha}(-\lambda+\alpha v^x-\beta^x), \\
{\cal B} &=& b^x - b^0\lambda, \\
G &=& \frac{1}{\alpha^2}(-(\lambda+\beta^x)^2+\alpha^2 \gamma^{xx}).
\end{eqnarray}

Let us note that in the previous discussion about the roots of the
characteristic polynomial we have omitted the fact that the entropy
waves as well as the Alfv\'en waves appear as double roots. These
superfluous eigenvalues appear associated with unphysical waves and are
the result of working with the unconstrained, $10 \times 10$ system of
equations. We note that \citet{vanputten91} derived a different augmented
system of RMHD equations in constrained-free form with different nonphysical
waves. Any attempt to develop a numerical procedure based on the wave
structure of the RMHD equations must remove these nonphysical waves
(and the corresponding eigenvectors) from the wave decomposition. In
the case of SRMHD \citet{komissarov99} and \citet{koldoba} eliminate 
the nonphysical eigenvectors by demanding the waves to preserve the values 
of the invariants $u^\mu u_\mu = -1$ and $u^\mu b_\mu = 0$ as suggested 
by \citet{anile}. Correspondingly, \citet{balsara01} selects the physical 
eigenvectors by comparing with the equivalent expressions in the 
nonrelativistic limit.

It is worth noticing that just as in the classical case, the relativistic 
MHD equations have degenerate states in which two or more wavespeeds 
coincide, which breaks the strict hyperbolicity of the system. 
\citet{komissarov99} has reviewed the properties of these degeneracies. 
In the fluid rest frame, the degeneracies in both classical and relativistic 
MHD are the same: either the slow and Alfv\'en waves have the same speed 
as the entropy wave when propagating perpendicularly to the magnetic field 
(Degeneracy I), or the slow or the fast wave (or both) have the same speed 
as the Alfv\'en wave when propagating in a direction aligned with the magnetic 
field (Degeneracy II). \citet{anton05} have characterized these degeneracies 
in terms of the components of the magnetic field four-vector normal and 
tangential to the Alfv\'en wavefront, ${\bf b}_n$, ${\bf b}_t$. When 
${\bf b}_n = 0$, the system falls within Degeneracy I, while Degeneracy 
II is reached when ${\bf b}_t = 0$. Let us note that the previous 
characterization is covariant (i.e.~defined in terms of four-vectors) and 
hence can be checked in any reference frame. In addition, \citet{anton05} 
have also worked out a single set of right and left eigenvectors which are 
regular and span a complete basis in any physical state, including degenerate 
states. The {\it renormalization} procedure can be understood as a relativistic
generalization of the work performed by \citet{brio} in classical
MHD. This procedure avoids the ambiguity inherent to a change of basis
when approaching a degeneracy, as done e.g.~by \citet{komissarov99}. The 
renormalized eigenvectors have been used in all the tests reported in the 
present paper using the {\it full-wave decomposition} Riemann solver.

%%%%%%%%%%%%%%%%%%%%%%%%%%%%%%%%%%%%%%%%%%%%%%%%%%
\section{Numerical Approach}
\label{III}
%%%%%%%%%%%%%%%%%%%%%%%%%%%%%%%%%%%%%%%%%%%%%%%%%%

Writing the GRMHD equations as a first-order, flux-conservative,
hyperbolic system allows us to use numerical methods specifically
designed to solve such kind of equations. Among these methods,
high-resolution shock-capturing (HRSC) schemes are recognized as the
most efficient schemes to evolve complex flows accurately, capturing
the discontinuities which appear when dealing with
nonlinear hyperbolic equations. 

%%%%%%%%%%%%%%%%%%%%%%%%%%%%%%%%%%%%%%%%%%%%%%%%%%%%%%%%%%%
\subsection{Integral form of the GRMHD equations}
%%%%%%%%%%%%%%%%%%%%%%%%%%%%%%%%%%%%%%%%%%%%%%%%%%%%%%%%%%%

To apply HRSC techniques to the present GRMHD system we use 
Eq.~(\ref{eq:fundsystem}) in integral form. Let $\Omega$ be a simply 
connected region of the four-dimensional manifold bounded by a closed 
three-dimensional surface $\partial\Omega$. We take $\partial\Omega$ as 
the standard-oriented hyperparallelepiped made up of the two spacelike
surfaces ${\Sigma_t, \Sigma_{t+\Delta t}}$ plus timelike surfaces
${\Sigma_{x^i},\Sigma_{x^i+\Delta x^i}}$, that connect the two temporal
slices. Then, the integral form of Eq.(\ref{eq:fundsystem}) is
\begin{equation}
\int_\Omega \frac{1}{\sqrt{-g}}
\frac {\partial \sqrt{\gamma}{\bf F}^{0}}
{\partial x^{0}} d\Omega +
\int_\Omega\frac{1}{\sqrt{-g}} \frac{\partial \sqrt{-g}{\bf F}^{i}}
{\partial x^{i}} d\Omega 
 = \int_\Omega {\bf S} d\Omega,
\end{equation}
which can be written, for numerical purposes, as follows
\vspace{1cm}
\begin{widetext}
\begin{eqnarray}
(\bar{\bf F}^{0})_{t+\Delta t}-(\bar{\bf F}^{0})_{t} &=&
-\left(\int_{\Sigma_{x^1+\Delta x^1}}\sqrt{-g}\hat{\bf F}^{1} dx^0 dx^2 dx^3 
      -\int_{\Sigma_{x^1}}           \sqrt{-g}\hat{\bf F}^{1} dx^0
      dx^2 dx^3\right) \nonumber \\
&& -\left(\int_{\Sigma_{x^2+\Delta x^2}}\sqrt{-g}\hat{\bf F}^{2} dx^0 dx^1 dx^3
      -\int_{\Sigma_{x^2}}           \sqrt{-g}\hat{\bf F}^{2} dx^0
      dx^1 dx^3\right) \nonumber \\
&& -\left(\int_{\Sigma_{x^3+\Delta x^3}}\sqrt{-g}\hat{\bf F}^{3} dx^0 dx^1 dx^2 
      -\int_{\Sigma_{x^3}}           \sqrt{-g}\hat{\bf F}^{3} dx^0
      dx^1 dx^2\right) + \int_\Omega {\bf S} d\Omega ,
\label{eq:system}
\end{eqnarray}
\end{widetext}
where
\begin{equation}
\bar{\bf F}^{0}=
\frac{1}{\Delta V}\int_{x^1}^{x^1+\Delta x^1} \int_{x^2}^{x^2+\Delta x^2} 
\int_{x^3}^{x^3+\Delta x^3} \sqrt{\gamma}{\bf F}^{0} dx^1dx^2dx^3 
\end{equation}
and 
\begin{equation}
\Delta V= \int_{x^1}^{x^1+\Delta x^1} \int_{x^2}^{x^2+\Delta x^2}
\int_{x^3}^{x^3+\Delta x^3} \sqrt{\gamma} dx^1dx^2dx^3.
\end{equation}
The carets appearing on the fluxes denote that these fluxes, which are
calculated at cell interfaces where the flow conditions can be
discontinuous, are obtained by solving Riemann problems between the
corresponding numerical cells. These numerical fluxes are further discussed
in Section~\ref{numflux}.

We note that in order to increase the spatial accuracy of the numerical solution,
the primitive variables (see Sect.~\ref{recovery}) are reconstructed at the cell
interfaces before the actual computation of the numerical fluxes. We use a 
standard second order {\it minmod} reconstruction procedure to compute the values 
of $p$, $\rho$, $v_i$ and $B^i$ ($i= 1,2,3$) at both sides of each numerical 
interface. However, when computing the numerical fluxes along a certain direction, 
we do not allow for discontinuities in the magnetic field component along that 
direction. Furthermore, the equations in integral form are advanced in time 
using the method of lines in conjunction with a second order, conservative 
Runge-Kutta method \citep{shu:88}.

%%%%%%%%%%%%%%%%%%%%%%%%%%%%%%%
\subsection{Induction equation}
%%%%%%%%%%%%%%%%%%%%%%%%%%%%%%%

The main advantage of the above numerical procedure, Eq.~(\ref{eq:system}), 
to advance in time the system of equations, is that those variables which 
obey a conservation law are, by construction, conserved during the evolution 
as long as the balance between the fluxes at the boundaries of the
computational  domain and the source terms are zero. This is an important 
property that any hydrodynamics code should fulfill. 

%However, the system of 
%equations~(\ref{eq:fundsystem}) also contains the components of the magnetic 
%field, which do not obey a conservation law but an induction equation, 
%Eq.~(\ref{evolB}). Therefore, the numerical advantage of using 
%Eq.~(\ref{eq:system}) for the conserved variables does not apply for the 
%magnetic field components. Moreover, the magnetic field must satisfy the
%divergence-free condition, Eq.~(\ref{divfree}), during the evolution. However,
%there is no guarantee that the divergence is conserved
%numerically when updating  the magnetic field if we were
%to use the same numerical procedure we employ for  the
%rest of components of the state vector. 
%-------------------------------------
% CHECK IF YOU AGREE WITH WHAT FOLLOWS
However, as far as the magnetic field components are
concerned, the system of 
equations~(\ref{eq:fundsystem}) only includes 
the induction equation Eq.~(\ref{evolB}), expressed by 
~(\ref{eq:fundsystem}) in conservation form, while the
divergence-free condition, Eq.~(\ref{divfree}), remains
as an additional constraint to be imposed.
Therefore, the numerical advantage of using 
Eq.~(\ref{eq:system}) for the conserved variables does
not apply straightforwardly
for the magnetic field components. 
Indeed, there is no guarantee that the divergence is conserved
numerically when updating  the magnetic field if we were
to use the same numerical procedure we employ for  the
rest of components of the state vector. 
%----------------------------------------

Among the methods designed to preserve the divergence of the magnetic field
we use the constrained transport method designed by
\citet{evans88} and first extended to HRSC methods by
\cite{ryu:98} 
(see also \cite{londrillo:04} for a recent discussion).
This scheme  is based on the use of
Stokes theorem after the integration of the induction
equation on surfaces of constant $t$ and $x^i$,
$\Sigma_{t,x^i}$. Let us write Eq.~(\ref{evolB}) as
\begin{equation}
\frac{1}{\sqrt{\gamma}}\frac{\partial {\vec{\cal B}}}{\partial
t}=\vec{\nabla} \times \vec{\Omega},
\label{omegaeq}
\end{equation}
where we have defined the density vector $\vec{\cal
B}=\sqrt{\gamma}\vec{B}$ and
$\vec{\Omega}=(\alpha\vec{v}-\vec\beta)\times\vec{B}$. 

To obtain a discretized version of Eq.~(\ref{omegaeq}), we proceed as follows.
At a given time, each numerical cell is bounded by 6 two-surfaces. Consider, for 
concreteness, the two-surface $\Sigma_{t,x^3}$, defined by $t={\rm const.}$~and 
$x^3={\rm const.}$, and the remaining two coordinates spanning the intervals from 
$x^1$ to $x^1+\Delta x^1$, and from $x^2$ to $x^2+\Delta x^2$. The magnetic flux  through this two-surface is given by 
\begin{equation}
\Phi_{\Sigma_{t,x^3}}=\int_{\Sigma_{t,x^3}} \vec{B} \cdot d\vec{\Sigma}.
\end{equation}
Furthermore, the electromotive force ${\cal E}$ around the contour 
$\partial(\Sigma_{t,x^3})$ is defined as 
\begin{equation}
{\cal E}(t)=-\int_{\partial(\Sigma_{t,x^3})} \Omega_i dx^i.
\end{equation}
Integrating Eq.~(\ref{omegaeq}) on the two-surface $\Sigma_{t,x^3}$, and 
applying Stokes theorem to the right hand side we obtain the equation
\begin{equation}
\frac{d\Phi_{\Sigma_{t,x^3}}}{dt}=-{\cal E}=\int_{\partial(\Sigma_{t,x^3})} 
\Omega_i dx^i,
\end{equation}
which can be integrated to give
\begin{equation}
\Phi^{t+\Delta t}_{\Sigma_{t,x^3}}-\Phi^{t}_{\Sigma_{t,x^3}}=
\int_t^{t+\Delta t} \int_{\partial(\Sigma_{t,x^3})}
\hat{\Omega}_i dx^i\;\; dt,
\label{eq:mflux}
\end{equation}
where the caret denotes again that quantities $\hat{\Omega}_i$ are calculated at the
edges of the numerical cells, where they can be discontinuous. At each edge, as we 
will describe below, these quantities are calculated using the solution of four Riemann
problems between the corresponding faces whose intersection defines the edge. However, 
irrespective of the expression we use for calculating $\hat{\Omega}_i$, the method to 
advance the magnetic fluxes at the faces of the numerical cells satisfies, by 
construction, the divergence constraint.  To see this we can integrate over a 
computational cell the divergence of the magnetic field at a given time. After applying 
Gauss theorem, we obtain
\begin{equation}
\int_{\Delta V} \nabla \cdot \vec{B} dV=\int_{\Sigma}\vec{B}\cdot 
d\vec{\Sigma}=\sum_{{\rm faces}, i=1}^6 \Phi_i.
\label{gauss}
\end{equation}
In the previous expression, $\Delta V$ stands for the volume of a
computational cell, whereas $\Sigma$ denotes the closed surface
bounding that cell. The summation is extended to the six faces
(coordinate surfaces) shaping $\Sigma$. Now, taking the time derivative
of Eq.~(\ref{gauss}) yields to
\begin{eqnarray}
\frac{d}{dt}\int_{\Delta V} \nabla \cdot \vec{B} dV&=&-\sum_{{\rm faces}, i=1}^6 
\frac{d}{dt}\Phi_i \nonumber \\
&=& \sum_{{\rm faces}, i=1}^6\sum_{{\rm edges}, j=1}^4 
{\cal E}_{ij},
\end{eqnarray}
where ${\cal E}_{ij}$ is the contribution from edge $j$ to the
total electromotive force around the contour defined by the boundary
of face $i$.  It turns out that the above summation cancels exactly
since the value of ${\cal E}$ for the common edge of two adjacent
faces has a different sign for each face. Therefore, if the initial
fluxes through each face of a numerical cell verify $\Sigma_{{\rm
faces}, i=1}^6\Phi_i=0$, this condition will be fulfilled during the
evolution. 

%%%%%%%%%%%%%%%%%%%%%%%%%%%%%%%%%%%%%%%%%%%%%%%%%%%%%%%%%%%
\subsection{Numerical fluxes and divergence-free condition}
\label{numflux}
%%%%%%%%%%%%%%%%%%%%%%%%%%%%%%%%%%%%%%%%%%%%%%%%%%%%%%%%%%%

The numerical integration of the GRMHD equations, Eqs.~(\ref{eq:fundsystem}) 
or (\ref{eq:system}), is done using a HRSC scheme. Such
schemes are specifically  designed to solve nonlinear
hyperbolic systems of conservation
laws~\citep{leveque,toro:97}. 
They are written in conservation form and use 
approximate or exact Riemann solvers to compute the numerical fluxes between
neighbour grid zones. This fact guarantees the proper capturing of all 
discontinuities which may arise naturally in the solution space of a nonlinear 
hyperbolic system. Applications of HRSC schemes in relativistic hydrodynamics 
can be found in \cite{martilr:03,fontlr}. 
Incidentally, we note that a detailed
description of linearized Riemann Solvers based on the spectral
decomposition can be found in \cite{font:94} for 
special relativistic hydrodynamics, and in 
\cite{banyuls:97} (diagonal metrics) and \cite{font:00},
\cite{ibanez:01}
(general metrics) for general relativistic hydrodynamics.
For HRSC methods in classical MHD, on the other hand, we
address to \cite{ryu:95,ryu:98}. 

As discussed in Section~\ref{II}, the existence of degeneracies in the 
eigenvectors of the RMHD system of equations makes it hazardous to implement
linearized Riemann solvers based on the full spectral decomposition of
the flux vector Jacobians. Nevertheless, we have succeeded in developing 
and implementing in the code a full-wave decomposition (Roe-type)
Riemann solver based on a single, renormalized set of right and left
eigenvectors, as discussed in detail in \citet{anton05}, which is
regular for any physical state, including degeneracies. This Riemann
solver is invoked in the code after a (local) linear coordinate
transformation based on the procedure developed by \citet{pons:98}
that allows to use special relativistic Riemann solvers in general
relativity, and which has been properly extended to include magnetic 
fields (see Sect.~\ref{SRRS}). 

In addition to the Roe-type Riemann solver we also use two simpler alternative
approaches to compute the numerical fluxes, namely the HLL single-state Riemann
solver of \citet{harten:83} and the second order central (symmetric) scheme
of \citet{tadmor} (KT hereafter). The KT scheme has proved recently to yield results
with an accuracy comparable to those provided by full-wave decomposition Riemann 
solvers in simulations involving purely hydrodynamical special relativistic 
flows~\citep{arturo} and general relativistic flows in dynamical neutron star 
spacetimes~\citep{shibata}. The interested reader is addressed to 
\citet{tadmor,arturo} for specific details on the KT central scheme.

Correspondingly, the HLL Riemann solver is based on the calculation of the 
maximum and the minimum left and right propagating wave
speeds emanating at the interface between the two 
initial states, and the resulting flux is
given by
\begin{eqnarray}
&& \hat{\bf F}({\bf U}_L,{\bf U}_R)= \nonumber \\
&& \frac{\tilde{\lambda}_{+}{\bf F}({\bf
U}_L)-\tilde{\lambda}_-{\bf F}({\bf U}_R) +
\tilde{\lambda}_ + \tilde{\lambda}_-  ({\bf U}_R-{\bf
U}_L)}{\tilde{\lambda}_+ + \tilde{\lambda}_-} \ ,
\end{eqnarray}
where $\tilde{\lambda}_{\pm}=\lambda_{\pm}/\alpha$.
Quantities $\hat{\bf F}$ stand for the numerical fluxes along each of
the three spatial coordinate directions, namely $\hat{\bf F}^i$ ($i=
1,2,3$) in Eq.~(\ref{eq:fundsystem}), whereas ${\bf U}\equiv{\bf F}^0$ denotes
the state vector. Subindices $L$ and $R$ stand for the left and right
states defining the Riemann problems at each numerical interface. Moreover
$\lambda_-$ and $\lambda_+$ are upper bounds of the speeds of the left- and 
right-propagating waves emanating from the cell interface,
\begin{eqnarray}
 \lambda_+ & = & \text{max}(0, \lambda^+_{{\rm fms},L},
 \lambda^+_{{\rm fms},R}),
 \\ 
 \lambda_- & = & \text{min}(0, \lambda^-_{{\rm fms},L},
 \lambda^-_{{\rm fms},R}),
\end{eqnarray}
where $\lambda^s_{{\rm fms},I}$ stands for the wavespeed of the fast
magnetosonic wave propagating to the left ($s= -$) or to the right ($s=+$)
computed at state $I$ ($=L,R$). These speeds are obtained
by looking for the smallest and largest solution of the
quartic equation ${\cal N}_4=0$ and can be effectively computed
with a Newton-Raphson iteration scheme
starting from $\lambda = \pm \alpha \sqrt{\gamma^{ii}}- \beta^i$ ($i= 1,2,3$).

Any of the flux formulae we have discussed can be used to advance the 
hydrodynamic variables according to Eq.~(\ref{eq:system}) and also to 
calculate the quantities $\hat{\Omega}_i$ needed to advance in time the 
magnetic fluxes following Eq.~(\ref{eq:mflux}). At each edge of the 
numerical cell, $\hat{\Omega}_i$ is written as an average of the numerical 
fluxes calculated at the interfaces between the faces whose intersection
define the edge. Let us consider, for illustrative purposes, $\hat{\Omega}_x$. 
If the indices $(j,k,l)$ denote the center of a numerical cell, an $x-$edge 
is defined by the indices $(j,k+1/2,l+1/2)$. By definition, 
$\Omega_x = \alpha(\tilde{v}^yB^z - \tilde{v}^zB^y)$. Since
\begin{equation}
\label{f1}
F^y(B^z) = \tilde{v}^yB^z - \tilde{v}^zB^y
\end{equation}
and 
\begin{equation}
\label{f2}
F^z(B^y) = \tilde{v}^zB^y - \tilde{v}^yB^z,
\end{equation}
we can express $\hat{\Omega}_x$ in terms of the fluxes as follows
\begin{eqnarray}
\label{oom}
\hat{\Omega}_{x\,j,k+1/2,l+1/2} &=& \frac{1}{4}
[\hat{F}^y_{j,k+1/2,l}+\hat{F}^y_{j,k+1/2,l+1} \nonumber \\
&& -\hat{F}^z_{j,k,l+1/2}-\hat{F}^z_{j,k+1,l+1/2}],
\end{eqnarray}
where $\hat{F}^y$($\hat{F}^z$) refers to the numerical flux in the $y$
($z$) direction corresponding to the equation for $B^z$ ($B^y$) and 
multiplied by $\alpha$ to account for the correct definition of $\Omega$. 
Also note that in the numerical implementation of the constraint transport
method, a slightly different procedure can be followed \citep{ryu:98}. 
According to this procedure, in the computation of the numerical fluxes 
(\ref{f1}) and (\ref{f2}), only the terms advecting the magnetic field
are considered (i.e.~the first term on the rhs of (\ref{f1})-(\ref{f2})), 
while the average in Eq.~(\ref{oom}) is obtained dividing by a factor 2 
instead of 4. Both of these procedures, the one described through
Eqs.~(\ref{f1})-(\ref{oom}) and its modification provided by \citet{ryu:98}
allow us to advance the magnetic flux at the faces of the numerical cells 
in the correct way. However, we have also noted that for 
2D numerical tests our implementation of this modified
scheme is generally more robust.
We address the interested reader to \cite{toth:00} for
additional properties of the \cite{ryu:98} scheme.

However, we need also to know the value of the magnetic field at the center 
of the cells in order to obtain the primitive variables after each time step 
(cf.~Sect.~\ref{recovery}) and to compute again the numerical fluxes of the 
other conserved variables for the next time step. If $\hat{B}^x_{j \pm 1/2,k,l}$ 
is the $x$-component of the magnetic field at the interface $(j\pm 1/2,k,l)$, 
then the $x$-component of the magnetic field at the center of the $(j,k,l)$ cell,
$B^x_{j,k,l}$, is obtained by taking the arithmetic average of the corresponding 
fluxes, i.e. 
\begin{eqnarray}
B^x_{j,k,l} =&& \frac{1}{2} (\hat{B^x}_{j-1/2,k,l}
\Delta S^x_{j-1/2,k,l}
+ \\ \nonumber
&&\hat{B^x}_{j+1/2,k,l}\Delta S^x_{j+1/2,k,l})/\Delta
S^x_{j,k,l} , \
\end{eqnarray}
where $\Delta S^x_{j\pm1/2,k,l}$ is the area of the interface surface between 
two adjacent cells, located at $x_{j\pm1/2}$ and bounded between 
$[y_{k-1/2},y_{k+1/2}]$ and $[z_{l-1/2},z_{l+1/2}]$. Analogous expressions for 
$\hat{\Omega}_{y\,j+1/2,k,l+1/2}$ and $\hat{\Omega}_{z\,j+1/2,k+1/2,l}$, and 
$B^y_{j,k,l}$ and $B^z_{j,k,l}$ can be easily derived.

%%%%%%%%%%%%%%%%%%%%%%%%%%%%%%%%%%%%%%%%%%%%%%%%%%%%%%%%%%
\subsection{Special relativistic Riemann solvers in GRMHD}
\label{SRRS}
%%%%%%%%%%%%%%%%%%%%%%%%%%%%%%%%%%%%%%%%%%%%%%%%%%%%%%%%%%

%% Inicio cambios Chema 3.5.05

  In \citet{pons:98} we presented a general procedure to use any
Riemann solver designed for the special relativistic hydrodynamics
equations in a general relativistic framework. In this section we
describe a generalization of this approach to account for the magnetic
field. It will be used to compute the numerical fluxes from the
special relativistic full-wave decomposition Riemann solver discussed
above. The procedure is based on performing linear transformations to
locally flat (or geodesic) systems of coordinates at each numerical
cell interface, from which the metric becomes locally Minkowskian
(plus second order terms). Notice that this approach is equivalent to
the usual approach in classical fluid dynamics where one uses the
solution of Riemann problems in slab symmetry for problems in
cylindrical or spherical coordinates. 
%Such an approximation breaks
%down near singular points (e.g.~the polar axis in cylindrical
%coordinates) and the spatial resolution would have to be increased
%accordingly.

  In order to generalize this procedure to the GRMHD case one must
start remembering that in the pure hydrodynamical case, the components
of the shift vector transversal to the cell interface play the role of
a {\it grid} velocity, i.e., as if we have a moving interface.  As
discussed in detail in \citet{pons:98}, this can be easily understood
by noticing that, the fluxes through the moving interface
for the local observer can be written as
$\bar{F}^i - \frac{\beta^i}{\alpha} F^0$, where $\bar{F}^i$ are the
fluxes when $\beta^i=0$ and $F^0$ the corresponding state vector. In
terms of $D$, $S_j$, $\tau$ and $p^{*}$, the structure of the first
five flux components (\ref{flux2}) in the magnetic case follow the
previous discussion with the conserved quantities advected with
$\tilde{v}^i$ (that includes the correction term for the moving grid)
and extra terms in the fluxes of momentum and energy (which do not
depend explicitly on the shift vector). This allows one to proceed
along the same steps as in \citet{pons:98}: i) Introduce the locally
Minkowskian coordinate system at each interface; ii) solve the Riemann
problem to obtain the numerical fluxes through the moving grid as seen
by the locally Minkowskian observer; iii) invert the transformation to
obtain the numerical fluxes in the original coordinates.

  Let us now concentrate in the last three components of the fluxes
(\ref{flux2}), namely $\tilde{v}^i B^k-\tilde{v}^k B^i$,
corresponding to the evolution of the magnetic
field. The terms $\tilde{v}^i B^k - v^k B^i$ also follow the
discussion for the non-magnetic case and the same numerical procedure
can be then applied. However, the term $\beta^k B^i/\alpha$ couples the
components of the shift vector parallel to the cell interface to the
perpendicular magnetic field.  This term has to be interpreted as a
correction to the total electromotive force caused by the movement of
the surface with respect to the Eulerian observer and has to be added
to the final expression for the flux.

  In Section~\ref{results} the validity of this approach with a
full-wave decomposition Roe-type Riemann solver is assessed in a
series of tests including discontinuous initial value problems, steady
flows, and dynamical accretion disks. As a result of this assessment
we conclude that the generalized procedure to use SR Riemann Solvers in
multidimensional GRMHD is an efficient and robust alternative to
develop specific solvers that need of the knowledge of the whole
spectral decomposition (eigenvalues and eigenvectors) in general
relativity.  Since each local change of coordinates is linear and it
only involves a few arithmetical operations, the additional
computational cost of the approach is negligible.

%%%%%%%%%%%%%%%%%%%%%%%%%%%%%%%%%%%%%%%%%%%%%%%%%%
\subsection{Primitive variable recovering}
\label{recovery}
%%%%%%%%%%%%%%%%%%%%%%%%%%%%%%%%%%%%%%%%%%%%%%%%%%

The numerical procedure used to solve the GRMHD equations allows us
to obtain the values of the conserved variables ${\bf F}^{0}$ at
time $t+\Delta t$ from their values at time $t$. However, the values
of the physical variables (i.e.~$\rho,\epsilon$, etc) are also needed 
at each time step in order to compute the fluxes. It is therefore 
necessary to solve the algebraic equations relating the conserved and 
the physical variables. For the classical MHD equations and an ideal 
gas equation of state the physical variables can be expressed as explicit
functions of the conserved ones. Unfortunately, this cannot be done in
GRMHD, a feature shared by the special and general relativistic versions 
of the purely hydrodynamics equations within the 3+1
approach (see 
\citet{papadopoulos} for an alternative formulation without this
shortcoming). Therefore, the resulting nonlinear algebraic system of equations 
has to be solved numerically. The procedure we describe below is an 
extension to full general relativity of that developed by \citet{komissarov99} 
in the special relativistic case.

The basic idea of this procedure relies on the fact that it is not
necessary to solve the system (\ref{conv_1})-(\ref{conv_3}) for the
three components of the momentum, but instead for its modulus $S^2
= S^iS_i$. The next step is to eliminate the components of
$b^\alpha$ through Eqs.~(\ref{b0})-(\ref{bi}). After some algebra it
is possible to write $S^2$ as
\begin{equation}
S^2 = (Z + B^2)^2 \frac{W^2-1}{W^2} - (2Z + B^2) \frac{(B^iS_i)^2}{Z^2},
\label{eq:s2}
\end{equation}
where $Z = \rho h W^2$.

The equation for the total energy can be worked out in a similar way
\begin{equation}
  \tau = Z + B^2 - p - \frac{B^2}{2W^2} - \frac{(B^iS_i)^2}{2Z^2} - D.
\label{eq:tau}
\end{equation}

Equations~(\ref{conv_1}), (\ref{eq:s2}) and (\ref{eq:tau}), together
with the definition of $Z$, form a system for the unknowns $\rho$, $p$
and $W$, assuming the function $h = h(\rho, p)$ is provided. In our
calculations we restrict ourselves to both, an ideal gas equation of
state (EOS), $p=\rho\epsilon(\gamma-1)$,
for which $h = 1 + \gamma p/\rho(\gamma-1)$, where $\gamma$ is
the adiabatic index, and a polytropic EOS (valid to describe isoentropic 
flows), $p=K\rho^\gamma$, where $K$ is the polytropic
constant. 
In this last case the integration of the total energy equation can be
avoided and the equation for the specific enthalpy is given by
\begin{equation}
  h = 1 + \frac{\gamma K}{\gamma - 1} \rho^{\gamma - 1}.
\end{equation}
Then Eqs.~(\ref{conv_1}) and 
(\ref{eq:s2}) are solved to obtain $\rho$ and $W$.

%%%%%%%%%%%%%%%%%%%%%%%%%%%%%%%%%%%%%%%%%%%%%%%%%%
\section{Results}
\label{results}
%%%%%%%%%%%%%%%%%%%%%%%%%%%%%%%%%%%%%%%%%%%%%%%%%%
%
\begin{figure*}
\begin{center}
\includegraphics[width=8.2cm,height=9.0cm,angle=0]{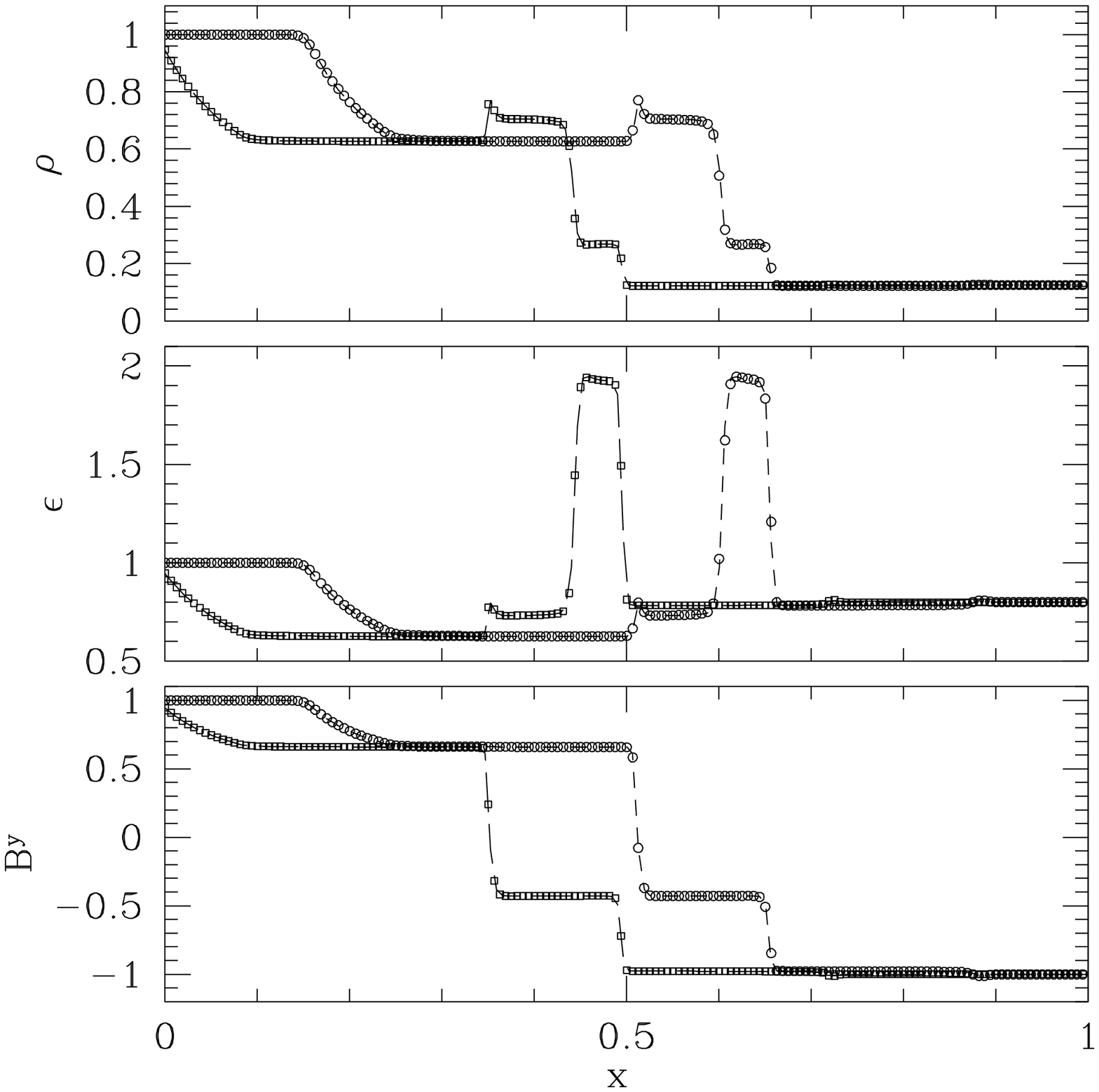}
\hspace{0.8truecm}
\includegraphics[width=8.2cm,height=9.0cm,angle=0]{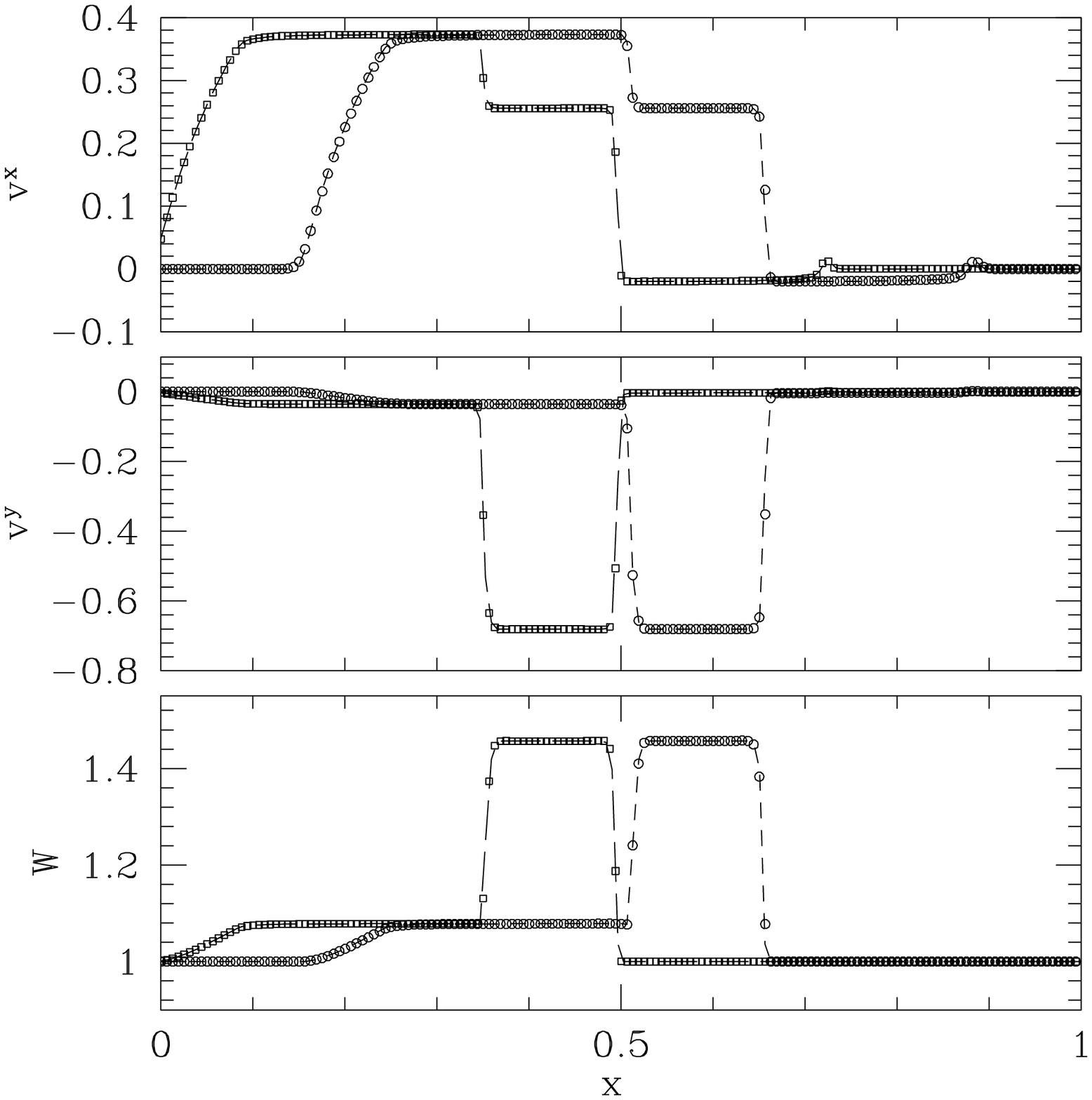}
\caption{ Wave pattern of the relativistic version of the Brio-Wu
shocktube test. The left panel reports the rest-mass density $\rho$,
the specific internal energy $\epsilon$ and the $y$ component of the
magnetic field $B^y$, while the right panel reports the $x$ and $y$
components of the velocity $v^x$ and $v^y$, and the Lorentz factor $W$.
The short dashed line shows the solution at time $t=0.4$ in 
Minkowski spacetime. The open circles represent the solution at time
$t=0.2$ in Minkowski spacetime with gauge effects mimicked by a
lapse function $\alpha=2.0$.  The open squares represent the solution
at time $t=0.4$ in Minkowski spacetime with a shift vector
$\beta^x=0.4$, while the long dashed line is the translation of the
short dashed one by the amount $\beta^x t=0.16$.  (Only 160 of the
1600 data points used in the simulation are drawn).  }
\label{fig1}
\end{center}
\end{figure*}

We turn now to assess the formulation of the GRMHD equations we have
presented as well as the numerical techniques we employ to solve them. 
The simulations reported in this section are introduced in a way which 
gradually increases the level of complexity of the flow to solve, 
starting first with shock tube tests in both purely Minkowski spacetime 
and flat spacetimes suitably modified by the presence of 
artificial gauge terms.
Next we turn to one-dimensional tests of accreting magnetized flows onto
Schwarzschild and Kerr black holes, to finally discuss two-dimensional
simulations of thick accretion disks orbiting around black holes. This
collection of tests allows us to validate our approach by comparing
the numerical simulations with analytic solutions (in the cases where
such a comparison is possible), by investigating the ability of the
code to preserve stationary solutions in the strong gravitational
field regime, and by comparing with available numerical results 
reported in the literature. 

For those tests which involve (background) black hole spacetimes we adopt 
Boyer-Lindquist coordinates and we fix the unit of length to 
$r_g\equiv M$, $M$ being the mass of the black hole.

%%%%%%%%%%%%%%%%%%%%%%%%%%%%%%%%%%%%%%%%%%%%%%%%%%
\subsection{Relativistic Brio-Wu shock tube test}
%%%%%%%%%%%%%%%%%%%%%%%%%%%%%%%%%%%%%%%%%%%%%%%%%%

The first test is the relativistic analog of the classical
Brio-Wu shock tube problem~\citep{brio,balsara01}, as adapted to the  
relativistic MHD case by \cite{vanputten93}. The computational setup
consists of two constant states which are initially at rest and
separated through a discontinuity
placed at the middle point of a unit length
domain.  The two states are characterized by the following initial
conditions: Left state: $\rho = 1.0$, $v^x = 0.0$, $v^y = 0.0$, $p =
1.0$, and $B^y = 1.0$. Right state: $\rho = 0.125$, $v^x = 0.0$, $v^y
= 0.0$, $p = 0.10$, $B^y = -1.0$. The adiabatic index of the ideal gas
EOS is $\gamma = 2 $, and the $x$ component of the magnetic field is 
equal for both left and right states, $B^x = 0.5$. The test is
performed using a Cartesian grid with 1600 cells. Results are reported
for the HLL Riemann solver (as the other two schemes yield similar
results) and for a CFL parameter equal to 0.5.

The results of the simulation are shown in Fig.~\ref{fig1}, which displays 
the wave structure for various quantities after the removal of the membrane.  
This wave structure comprises a fast rarefaction wave, a slow compound wave 
(both moving to the left), a contact discontinuity, and, moving to the
right, a slow shock wave and a fast rarefaction wave. The short dashed line 
in the six panels of Fig.~\ref{fig1} shows the wave pattern produced in the 
purely Minkowski spacetime at time $t=0.4$. It is in good overall agreement 
with the results obtained by~\citet{balsara01}, in particular regarding the 
location of the different waves, the maximum value achieved by the Lorentz 
factor ($W=1.457$), and the smearing of the numerical solution. In addition 
to this solution we use open circles to denote the results of this test in 
flat spacetime but incorporating {\it gauge} effects by selecting a value 
of the lapse function different from unity, namely $\alpha=2$. The solution, 
which is shown at $t=0.2$, matches as expected with that represented by the 
short dashed line, obtained in flat spacetime at time $t=0.4$.  Finally, 
the open squares refer to a third version of this test carried out in a flat
spacetime with a nonvanishing shift vector, namely $\beta^x=0.4$. The numerical
displacement that is thus produced is in perfect agreement with the expected 
one. This is emphasized in the figure by translating the short dashed line into 
the long dashed one by the predicted amount, $\beta^x t=0.16$.

%%%%%%%%%%%%%%%%%%%%%%%%%%%%%%%%%%%%%%%%%%%%%%%%%%
\subsection{Magnetized spherical accretion}
\label{michel}
%%%%%%%%%%%%%%%%%%%%%%%%%%%%%%%%%%%%%%%%%%%%%%%%%%

\begin{figure*}
\begin{center}
\includegraphics[width=8.2cm,angle=0]{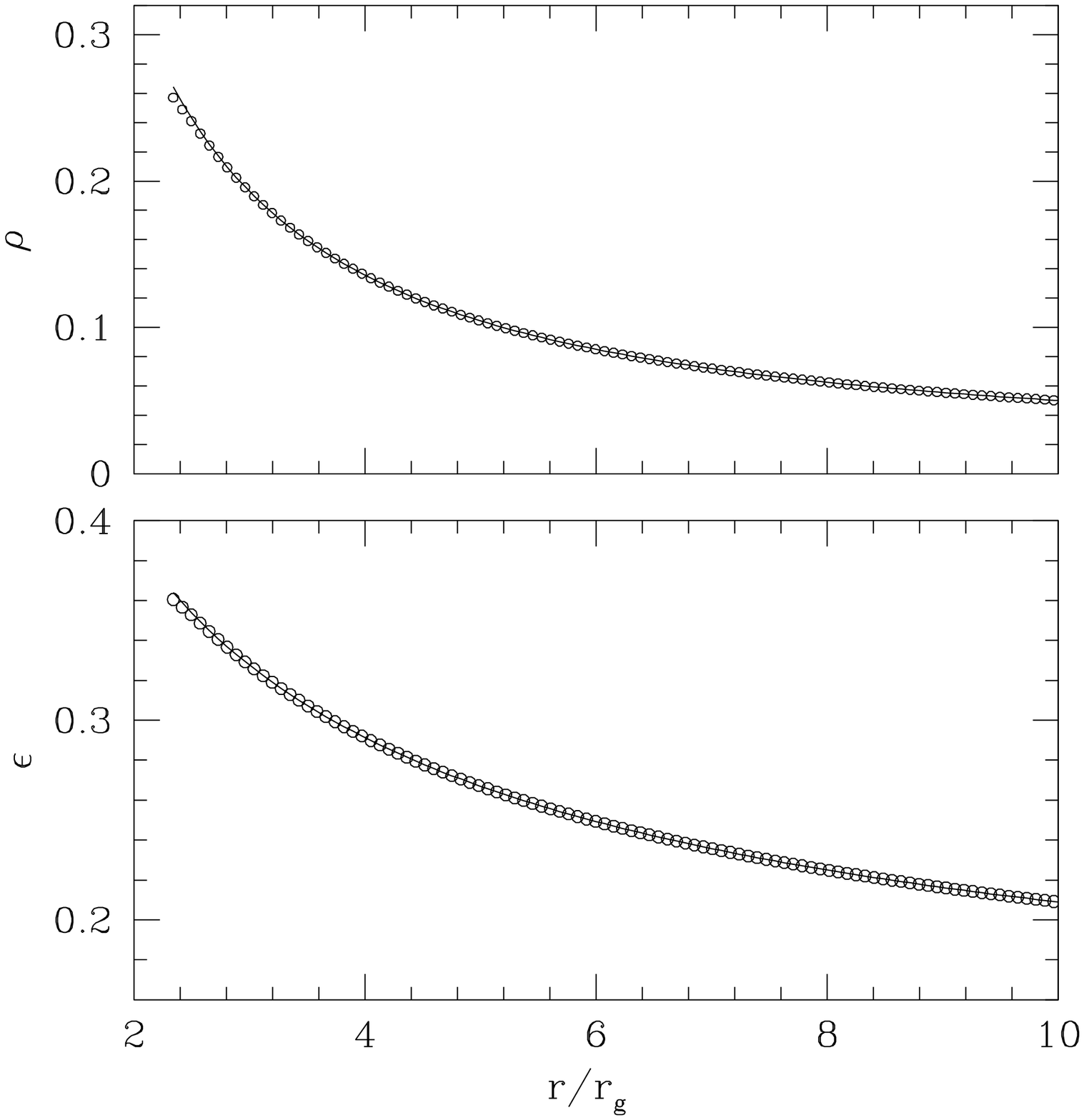}
\hspace{0.8truecm}
\includegraphics[width=8.2cm,angle=0]{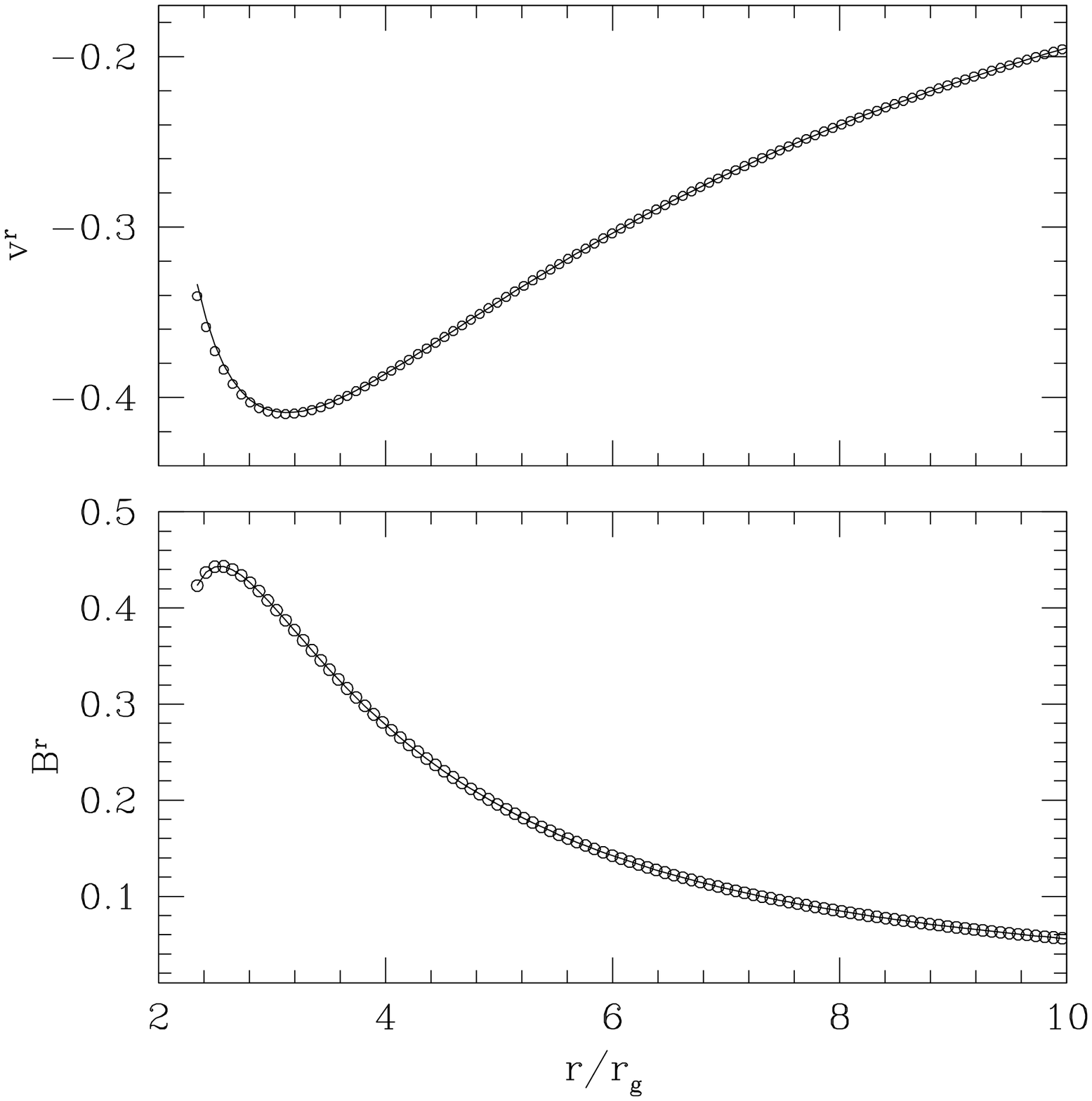}
\caption{Magnetized spherical accretion.  
Comparison between the analytic 
solution (solid line) and the numerical solution obtained
with the Roe-type Riemann Solver (circles) for a model with 
$\beta=1.0$ and $N=100$, once convergence is reached. 
The left panels display 
the radial profiles of the rest mass density $\rho$ (top) and the specific 
internal energy $\epsilon$, while the right panels show the corresponding 
profiles for the coordinate velocity $v^r$ (top) and the radial component 
of the magnetic field $B^r$, all of them in geometrized units.}
\label{fig2}
\end{center}
\end{figure*}

In the second test we check the ability of the code to numerically maintain
with a time-dependent system of equations the stationarity of the spherically 
symmetric accretion solution of a perfect fluid onto a Schwarzschild black 
hole in the presence of a radial magnetic field. It is worth emphasizing that 
a consistent solution for magnetized spherical accretion with a force-free
magnetic field satisfying the whole set of Maxwell equations does not exist 
(see Appendix~\ref{app_A} for a proof). However, it is easy to show that any magnetic 
field of the type $b^\alpha=(b^t,b^r,0,0)$ does not affect the spherically 
symmetric hydrodynamical solution. Therefore, although the resulting 
configuration is nonphysical, it provides a useful numerical test and has 
been used in the literature for this purpose \citep{gammie:03,devilliers1,duez05}.

The initial setup consists of a perfect isoentropic fluid obeying a polytropic 
EOS with $\gamma=4/3$. The critical radius of the solution is located at
$r_c=8.0$ and the rest mass density at the critical radius is $\rho_c=6.25
\times 10^{-2}$.  These parameters suffice to provide the full description 
of the spherical accretion onto a Schwarzschild black hole as described in 
detail by \citet{michel:72}. The radial magnetic field component, which can 
in principle follow any radial dependence, is chosen to satisfy the
divergence-free condition. Moreover, its strength is characterized by the
ratio $\beta=b^2/2p$ between the magnetic pressure and the gas pressure,
computed at the critical radius of the flow. These initial conditions are 
evolved in time using the Roe-type Riemann solver
described in Sec.~\ref{SRRS} on a uniform radial grid 
covering the region between $r_{\rm {min}}=r_{\rm {horizon}}+ \delta$ 
and $r_{\rm {max}} =10.0$, where $\delta$ varies from $0.1$ to $0.3$.  

\begin{figure}
\centerline{\psfig{file=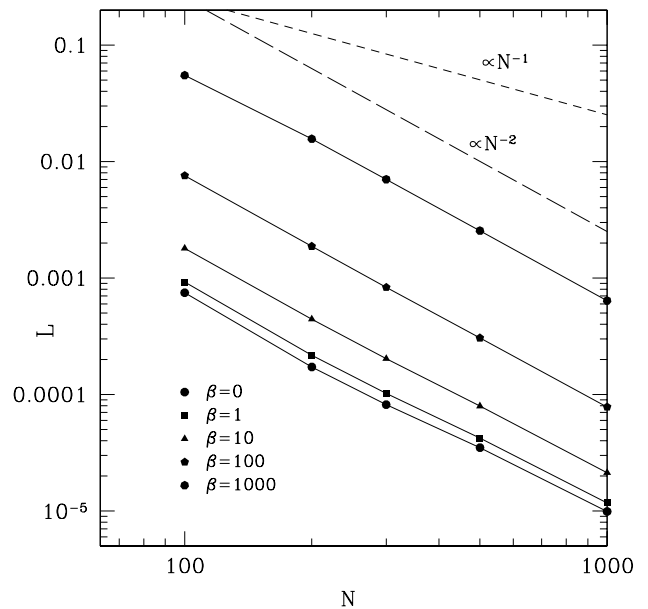,angle=0,width=8.5cm}}
\caption{
\label{fig3}
Error $\rm{L}$ of the rest mass density (see text for definition) for the 
magnetized spherical accretion test when the grid resolution is increased. 
The short-dashed and long-dashed lines indicate first and second order of 
global convergence, respectively. The symbols denote different values of 
the magnetization parameter at the critical point.}
\end{figure}

Figure~\ref{fig2} shows the comparison between the analytic solution (solid 
lines) and the numerical solution (circles) for one representative case with 
pressure ratio $\beta=1.0$ and $\delta=0.3$. These results are obtained with 
a numerical grid of $N=100$ radial zones, for which convergence is reached 
at time $t=250 M$. The order of accuracy of the code is computed by monitoring 
the error ${\rm L}\equiv\sum_{i=1}^N|Q_i - Q_{a,i}|/\sum_{i=1}^N Q_{a,i}$ for quantity 
$Q=\rho$ as the number of grid points $N$ is increased, where $Q_a$ 
represents the analytic solution. This procedure is repeated for different 
values of the ratio $\beta$, namely for $\beta=0, 1, 10, 100$, and $1000$ and 
the results, which are reported in Fig.~\ref{fig3}, show that the global order 
of convergence of the code is $2$, irrespective of the parameter $\beta$.

A comparison of the accuracy of the three methods we use to compute the
numerical fluxes is reported in Table~\ref{table1}, for $\beta=10.0$ and
$N=70$ radial zones. The results for the magnetized spherical accretion 
test appear in the upper half of the table. This table reports the global 
error of some representative quantities when numerical
convergence is reached.
For the particular test discussed in this section we find 
that there is not a single method providing the smallest
error in all of the quantities, and  the Roe-type
Solver, which is the most accurate in the computation of
the hydrodynamic variables, is the least accurate in the
computation of the magnetic field.
%$\delta Q_{\rm KT}<\delta Q_{\rm ROE}<\delta Q_{\rm HLL}$.

%%%%%%%%%%%%%%%%%%%%%%%%%%%%%%%%%%%%%%%%%%%%%%%%%%
\subsection{Equatorial Kerr accretion}
%%%%%%%%%%%%%%%%%%%%%%%%%%%%%%%%%%%%%%%%%%%%%%%%%%

\begin{figure*}
\begin{center}
\includegraphics[width=8.2cm,angle=0]{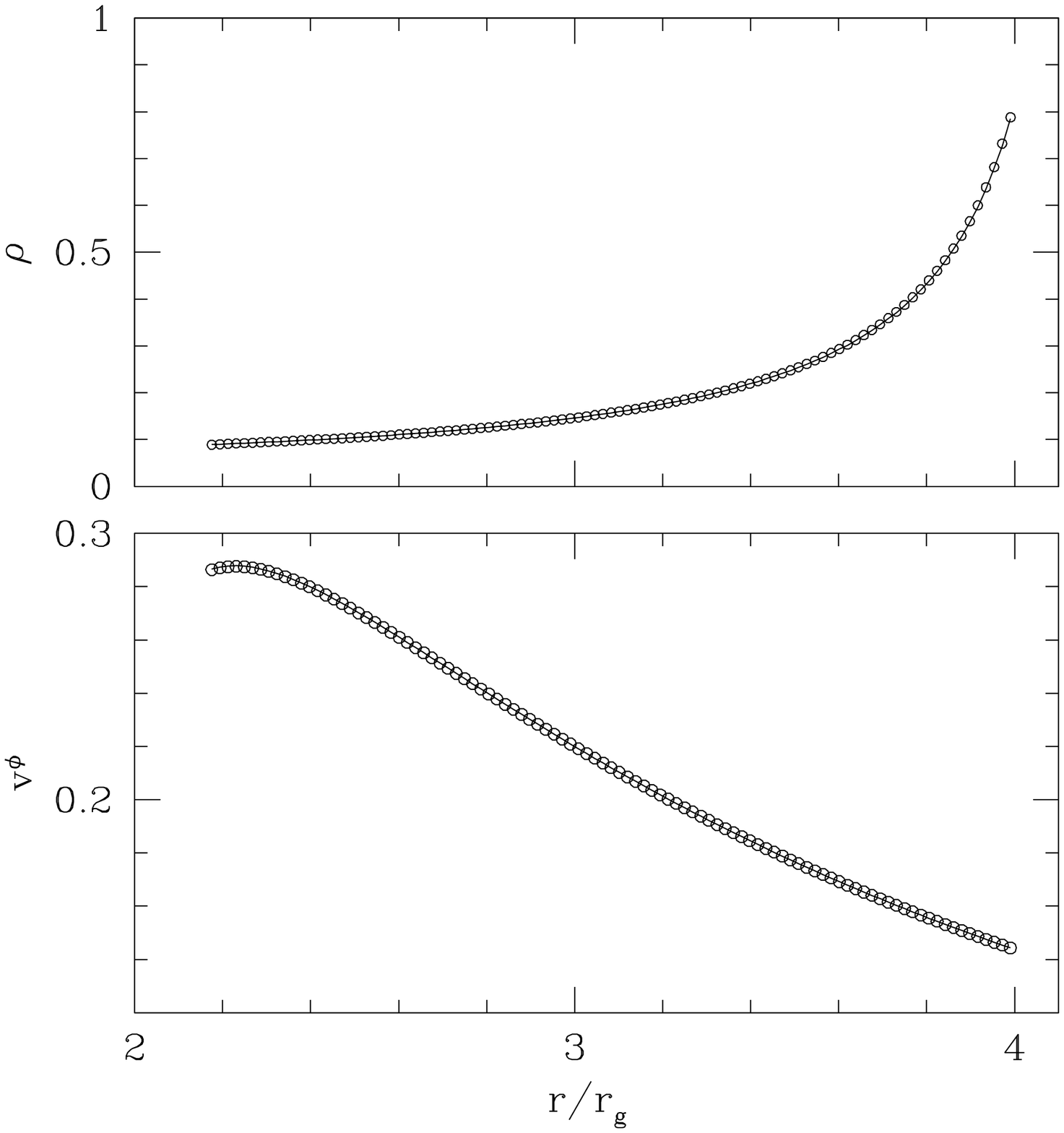}
\hspace{0.8truecm}
\includegraphics[width=8.2cm,angle=0]{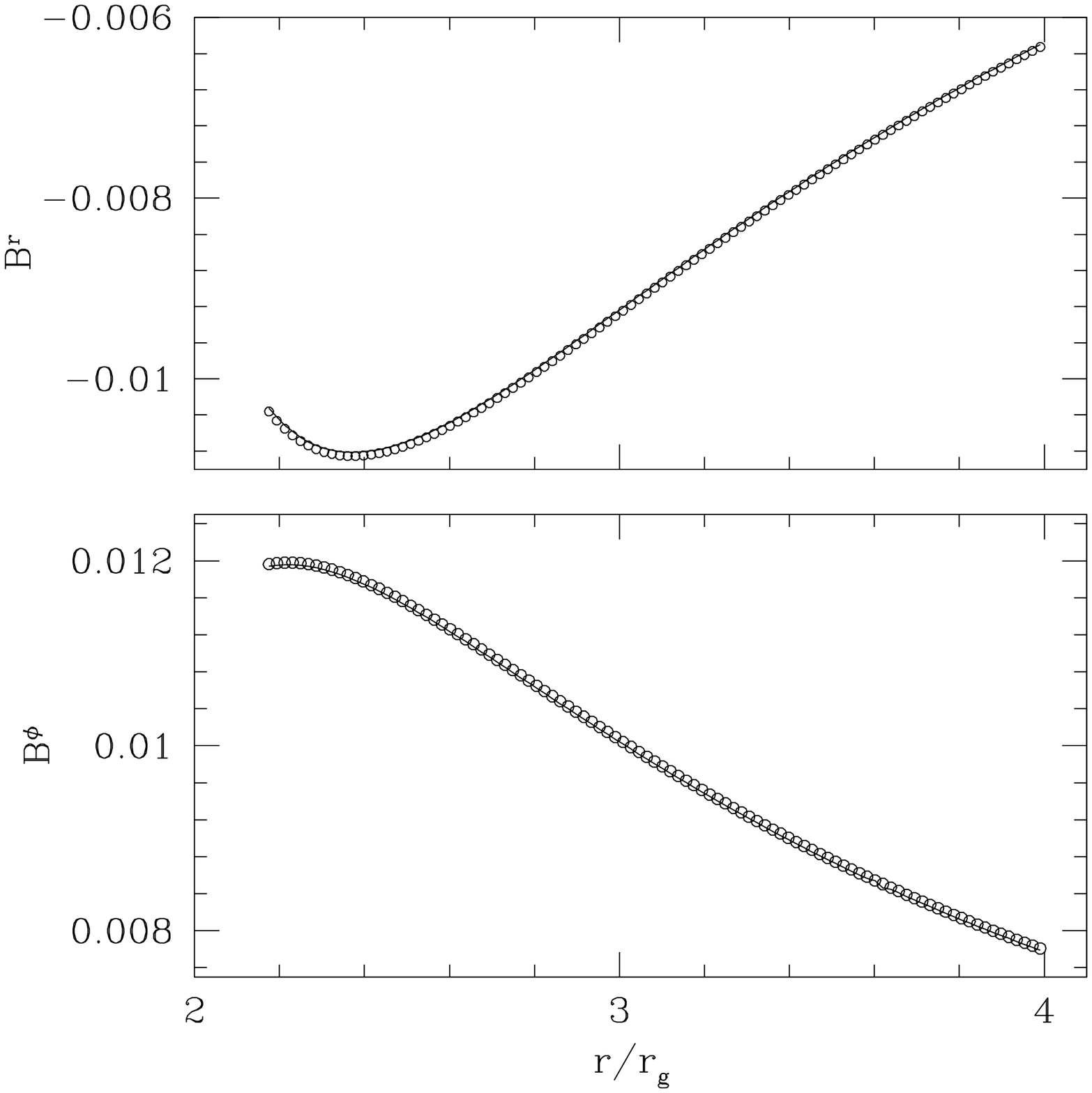}
\caption{ 
Comparison between the analytic 
solution (solid line) and the numerical converged 
solution obtained with the Roe-type Riemann Solver
(circles) for the magnetized
accretion solution onto a Kerr black hole with spin parameter $a=0.5$.
The left panel reports the rest mass density $\rho$ and the azimuthal
velocity $v^\phi$, while the right panel reports the radial and the
azimuthal components of the magnetic field, $B^r$ and $B^\phi$, all of
them in geometrized units.  }
\label{fig4}
\end{center}
\end{figure*}

A further one-dimensional test of the code is provided by
the stationary magnetized  inflow solution in the Kerr
metric derived by~\citet{takahashi}. This solution was
subsequently adapted to the case of equatorial inflow in
the region between the black  hole horizon and the
marginally stable orbit by~\cite{gammie}. This test has
been used  by~\citet{devilliers1} and ~\citet{gammie:03}
in the validation of their GRMHD codes.  It represents a
step forward in the level of complexity of the equations
to solve  with respect to those used in the previous two
sections, since the test involves the  Kerr metric,
albeit specialized to the equatorial plane. As a result,
additional  terms due to the increased number of
nonvanishing Christoffel symbols appear in the  equations.

\begin{figure}
\centerline{
\psfig{file=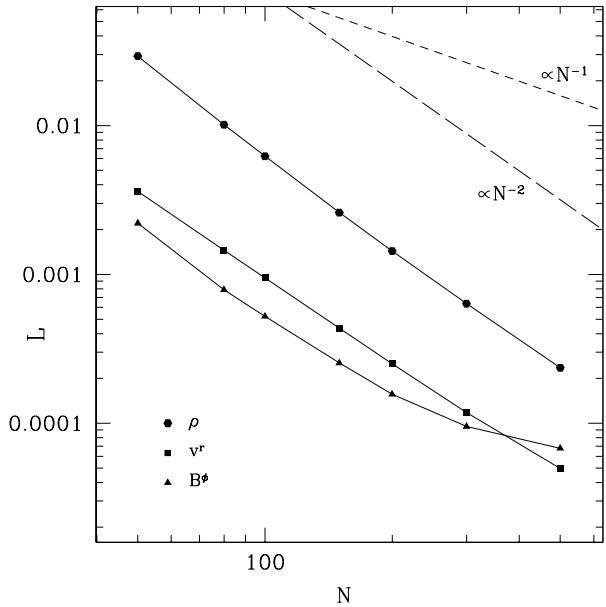,angle=0,width=8.5cm}
        }
\caption{
\label{fig5}
Error $\rm{L}$ of the rest mass density, the radial velocity and the
toroidal magnetic field for the magnetized equatorial accretion in the
Kerr metric.  The dashed and the long-dashed line indicate first and
second order of global convergence, respectively.  }
\end{figure}

As described by \citet{gammie} and adopting his notation,
the inflow solution is  determined once four conserved
quantities are specified, namely the accretion rate
$F_M$, the angular momentum flux $F_L$, the energy flux
$F_E$ and the component  $F_{\theta\phi}$ of the
electromagnetic tensor, which is related to the magnetic
flux through the inner edge of the disk. For the sake of
comparison we consider  an initial setup with the same
numerical values used by \citet{gammie:03}, namely  a
Kerr black hole with spin parameter $a=0.5$, $F_M=-1.0$,
$F_L=-2.815344$, $F_E=-0.908382$, $F_{\theta\phi}=0.5$.

The numerical grid consists of $N_r\times N_\theta$
gridpoints in the radial and  angular directions,
respectively. The radial grid covers the region between
$r_{\rm min}=r_{\rm horizon} + 0.2 $ and $r_{\max}=4.0$,
while the angular grid  consists of $N_\theta=3$
gridpoints subtending a small angle of $10^{-5}\pi$
accross the equatorial plane. The radial profiles of some
significant variables,  obtained with the Roe-type Riemann
solver, are reported in Fig.~\ref{fig4}
for a radial grid of $N_r=100$ zones. The open  circles
indicate the numerical results 
while the underlying solid lines correspond  to the
analytic solution. It is found that the stationarity of
the solution is  preserved to high accuracy by the
numerical code. For the long-term evolutions  considered
there are no significant deviations from the analytic
profiles. 

As we did for the magnetized spherical accretion test we
use the current test to compute again the order of
convergence of the code as the grid is refined. The
global order of convergence for some representative
quantities is reported in  Fig.~\ref{fig5}, which shows
that the code is second order accurate. As already
commented by \cite{gammie:03}, the worsening of the order
of convergence for  $B^{\phi}$ at high grid resolution is
due to the fact that the initial condition  is
``semi-analytic'', requiring the solution of an algebraic
equation. Thus, the inaccuracies produced at time $t=0$
become more pronunced for large numbers of radial zones $N_r$.

The performance of the code using the HLL and KT
solvers has also been checked with this test. While the
order of convergence is preserved irrespective  of the
numerical shemes used to compute the fluxes, the actual
accuracy can vary  significantly.  The results of this
comparison for the equatorial Kerr accretion  solution
are summarized in the lower half of Table~\ref{table1},
which reports the  global error of representative
quantities, when convergence is reached, on a numerical
grid with $N_r=60$ radial points. 
It is worth stressing that the HLL scheme, at least
in our implementation, turns out to be  the most accurate
in the computation of the magnetic field. 

\begin{widetext}
\begin{center}
\begin{deluxetable}{lcccc}
\tablecaption{Comparison among different schemes\label{table1}}
\tablehead{
\colhead{}
& \colhead{$\delta\rho$}
& \colhead{$\delta v^r$}
& \colhead{$\delta B^r$}
& \colhead{$\delta B^\phi$}
}
\startdata
Michel test & & & &  \\ \hline 
HLL   & $3.76\times 10^{-3}$ & $3.92\times 10^{-3}$ & $7.64\times 10^{-17}$  &  $-$    \\
Roe-type   & $2.97\times 10^{-3}$ & $3.45\times 10^{-3}$ & $1.09\times 10^{-12}$  &  $-$    \\
KT   & $3.36\times 10^{-3}$ & $3.54\times 10^{-3}$ & $1.94\times 10^{-18}$  &  $-$    \\
\hline 
Gammie test & & & & \\ \hline 
HLL   & $1.92\times 10^{-2}$ & $2.54\times 10^{-3}$ &
 $2.28\times 10^{-9}$  &  $1.48\times 10^{-3} $    \\
Roe-type   & $6.90\times 10^{-3}$ & $3.01\times 10^{-3}$ &
 $3.96\times 10^{-3}$  &  $2.14\times 10^{-3} $    \\
KT   & $1.63\times 10^{-2}$ & $9.72\times 10^{-4}$ &
 $2.30\times 10^{-9}$  &  $9.89\times 10^{-3} $    \\

\enddata
%\hline 
\tablecomments{Accretion tests: Comparison of
the accuracy of some representative  quantities for the HLL,
Roe and KT solvers. The columns report the global  errors
when convergence is reached.}
\end{deluxetable}
\end{center}
\end{widetext}
%\end{table}

%%%%%%%%%%%%%%%%%%%%%%%%%%%%%%%%%%%%%%%%%%%%%%%%%%%%%
\subsection{Thick accretion disks around black holes}
%%%%%%%%%%%%%%%%%%%%%%%%%%%%%%%%%%%%%%%%%%%%%%%%%%%%%

An intrinsic two-dimensional test for the code is provided by the stationary 
solution of a thick disk (or torus) orbiting around a black hole, described by 
\citet{fish:76}, \citet{kow:78}, and more recently by \citet{font:02a}. The 
resulting configuration consists of a perfect barotropic fluid in circular 
non-Keplerian motion around a Schwarzschild or Kerr black hole, with pressure 
gradients in the vertical direction accounting for the disk thickness. These 
thick disks may posses a cusp on the equatorial plane through which matter can
accrete onto the black hole.  

In the following two subsections we describe our numerical tests for unmagnetized 
and magnetized thick disks, respectively. In both cases the effective potential at 
the inner edge of the disk is smaller than that at the cusp, thus providing initial 
conditions which are strictly stationary. For simplicity we limit our simulations 
to models with constant distribution of specific angular momentum $\ell=-u_\phi/u_t$,
although the same qualitative results have been obtained with more general rotation 
laws.

%%%%%%%%%%%%%%%%%%%%%%%%%%%%%%%%%
\subsubsection{Unmagnetized disk}
\label{hydro_torus}
%%%%%%%%%%%%%%%%%%%%%%%%%%%%%%%%%

In testing the evolution of a purely hydrodynamical torus we consider a model 
similar to the one used by \cite{devilliers1} for the Schwarzschild metric, namely 
a torus with specific angular momentum $\ell=4.5$, position of the maximum density 
at $r_{\rm center}=15.3$, and an effective potential at the inner edge such that 
the inner and outer radii on the equatorial plane are $r_{\rm in}=9.34$ and
$r_{\rm out}=39.52$, respectively. We choose a polytropic EOS with $\gamma=4/3$ 
and a polytropic constant $K$ such that the torus-to-hole mass ratio is $M_t/M\sim 
0.07$.  

\begin{figure}
\centerline{
\psfig{file=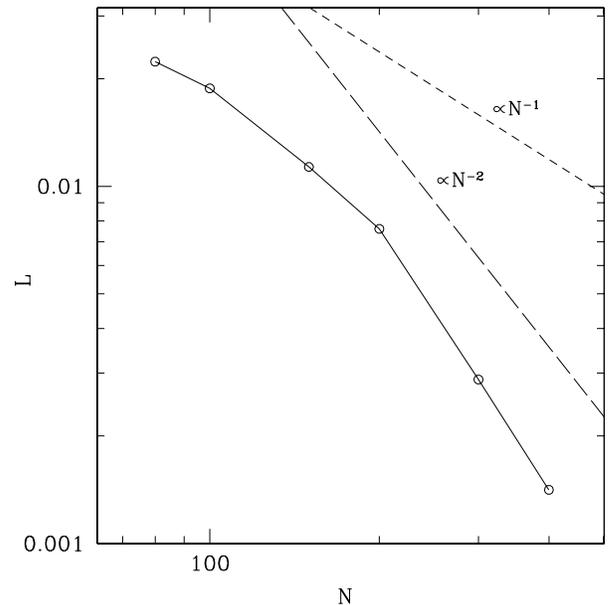,angle=0,width=8.5cm}
        }
\caption{
\label{fig6}
Unmagnetized thick accretion disk. Error $\rm{L}$ of the rest mass density when
resolution is increased. The short-dashed and the long-dashed lines indicate
first and second order of global convergence, respectively.  }
\end{figure}

We have checked that the code can keep the stationarity of the initial equilibrium 
torus when evolved in time. Figure \ref{fig6} shows the global order of convergence 
as computed from the rest mass density $\rho$.  The corresponding global error 
$\rm{L}$ reported in the figure, and defined as ${\rm L}\equiv\sum_{i,j=1}^N|\rho_{ij} - 
\rho_{a,ij}|/\sum_{i,j=1}^N\rho_{a,ij}$, is computed after $10$ orbital periods for each model, 
using a uniform numerical grid consisting of $N\times N$ gridpoints, whose specific
values can be read off from the figure.  As it is apparent from Fig.~\ref{fig6} the 
code reaches second order of convergence for reasonable high values of $N$ ($>200$).

\begin{figure*}
\begin{center}
\hspace{0.001cm}
\psfig{file=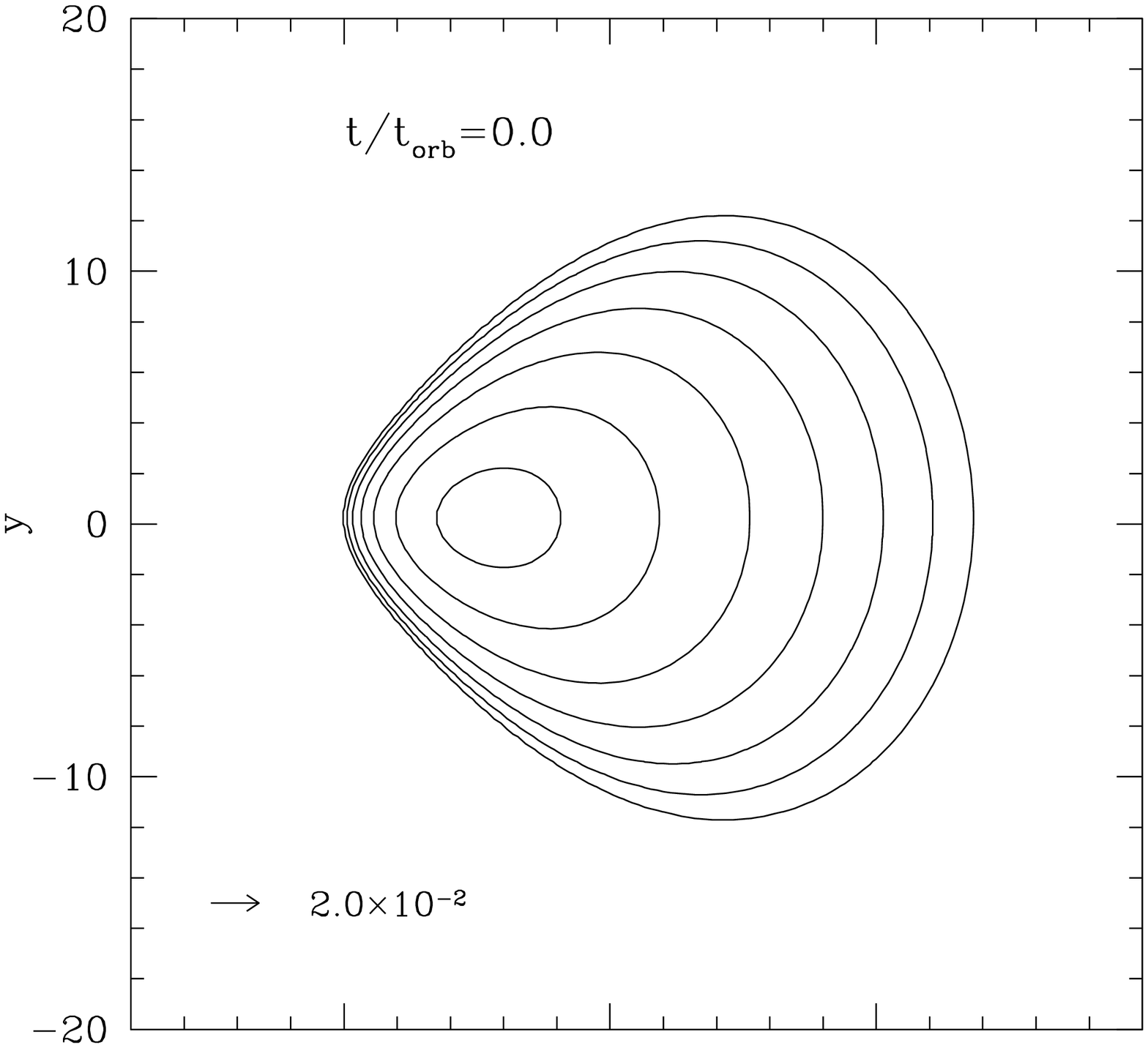,angle=0,width=7.5cm,height=7.5cm}
\hspace{-1.400cm}
\psfig{file=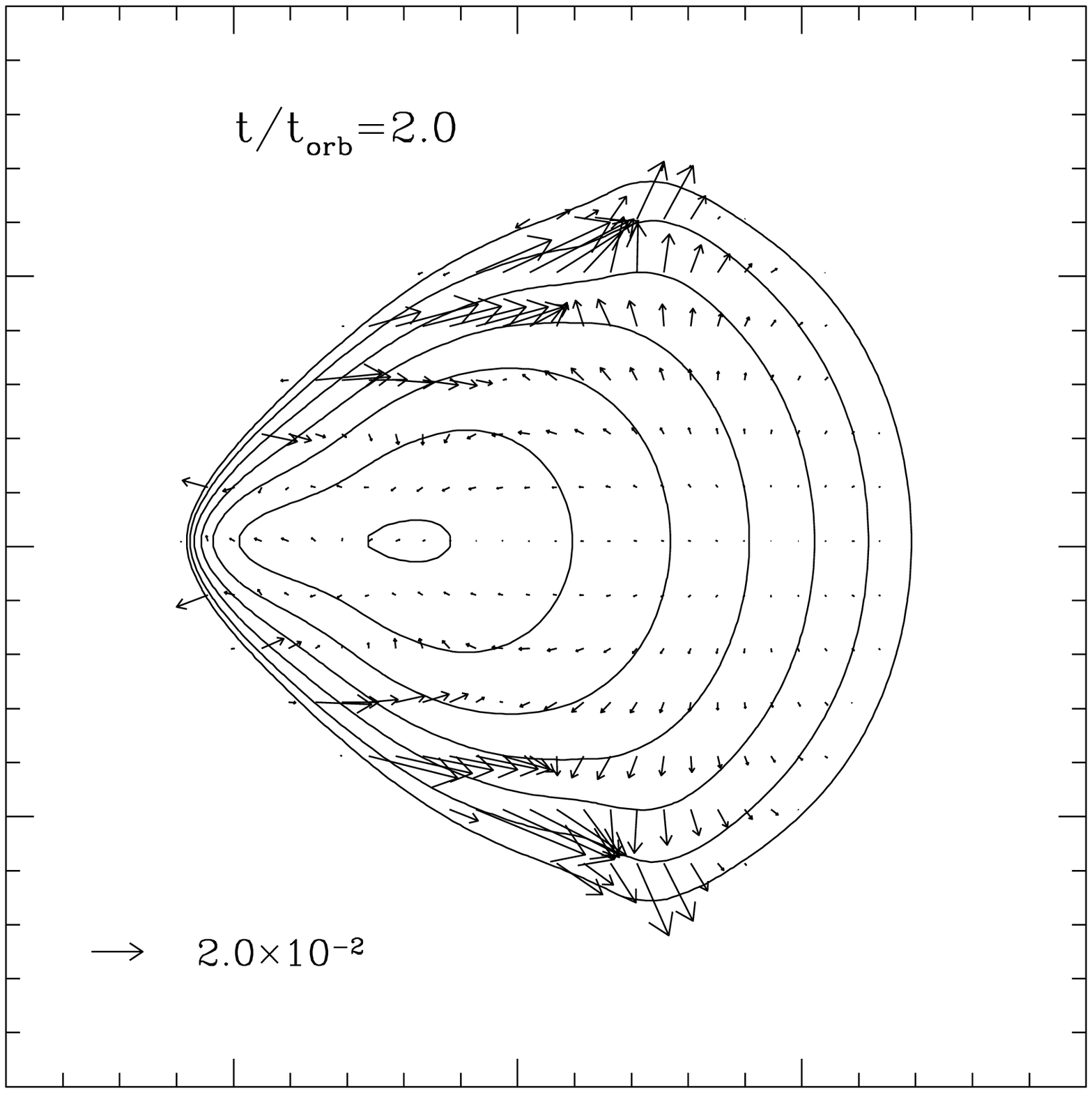,angle=0,width=7.5cm,height=7.5cm}
\vspace{-1.250cm}
\hspace{0.001cm}
\vspace{-1.250cm}
\psfig{file=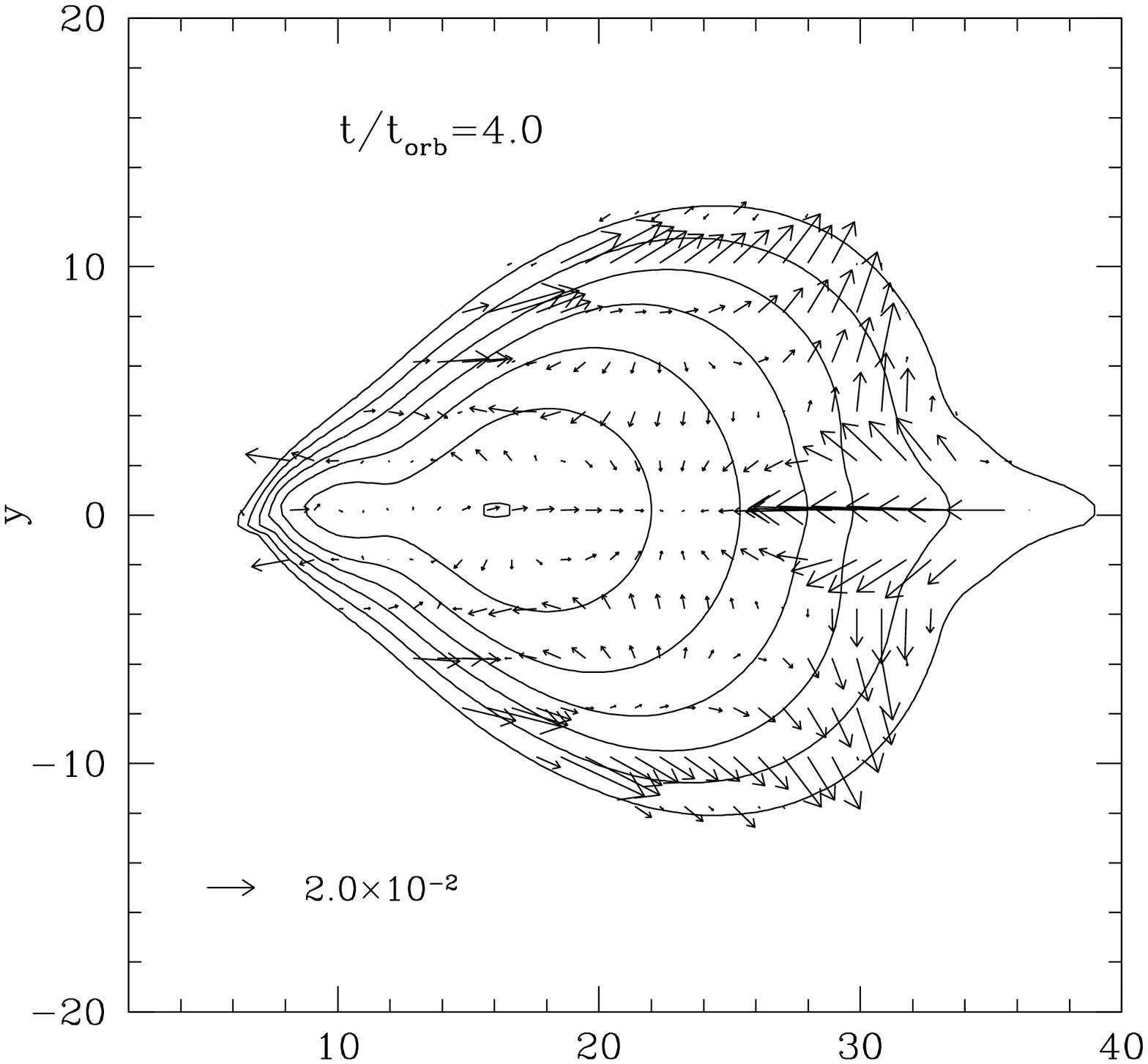,angle=0,width=7.5cm,height=7.5cm}
\hspace{-1.400cm}
\psfig{file=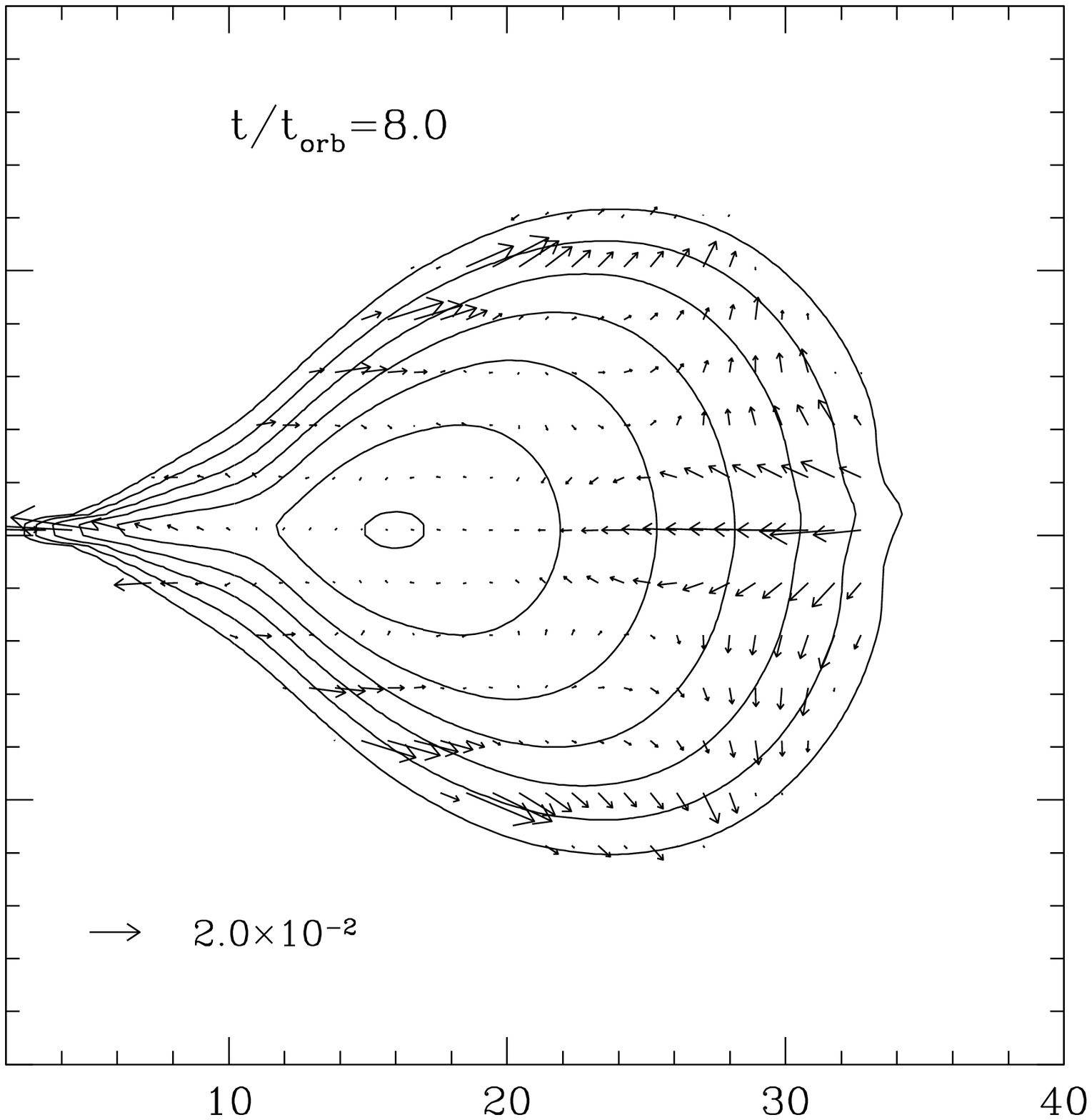,angle=0,width=7.5cm,height=7.5cm}
\end{center}
\vspace*{0.5cm}
\caption{Velocity field and logarithimic isocontours of
the rest mass density. At four orbital periods an
elongated high density structure is formed near the
equatorial plane, signaling the MRI instability in
2D calculations.}
\label{fig8}
\end{figure*}
We note that in addition to the model just discussed we have also analyzed the
performance of the code by comparing the evolution of additional hydrodynamical 
models which were studied by \citet{font:02a} and \citet{zanotti:03} using 
independent codes based on HRSC schemes. In all the cases considered, corresponding 
to a number of different generalizations such as disks with power-law distributions 
of the specific angular momentum, disks in Kerr spacetime, and disks subject to the 
so-called runaway instability, the GRMHD code reproduced the same quantitative 
results of the independent hydrodynamical codes with negligible differences.

%%%%%%%%%%%%%%%%%%%%%%%%%%%%%%%%%
\subsubsection{Magnetized disk}
\label{mag_torus}
%%%%%%%%%%%%%%%%%%%%%%%%%%%%%%%%%

As a final test we consider the evolution of a magnetized torus around a 
Schwarzschild black hole. In this case, however, a stationary solution 
which might provide self-consistent initial data for such
magnetized disks is not available. Indeed, 
it can be proved (see Appendix~\ref{app_B} for a proof) that the hydrodynamical
isoentropic type of models that we have used in the previous 
section for unmagnetized disks cannot be ``dressed'' with a magnetic field, 
to produce a force-free magnetized torus that satisfies the whole set of
Maxwell's equations. Therefore, we follow the same pragmatic approach adopted 
by \citet{devilliers1} and \citet{gammie:03}, and simply add an ad-hoc poloidal 
magnetic field to the hydrodynamical thick disk model. The magnetic field is 
generated by a vector potential $A_\phi\propto
\max(\rho/\rho_c - C, 0)$, where  $\rho_c$ is the maximum
rest mass density of the torus and $C$ is a free
parameter which determines the confinement of the field
inside the torus. The hydrodynamical
torus is the same as the one considered in
Section~\ref{hydro_torus}, but endowed with a magnetic
field characterized by a confinement parameter $C=0.5$
and such that the average ratio of magnetic-to-gas
pressure inside the torus is $\beta=1.5\times 10^{-3}$.  

\begin{figure}
\centerline{\psfig{file=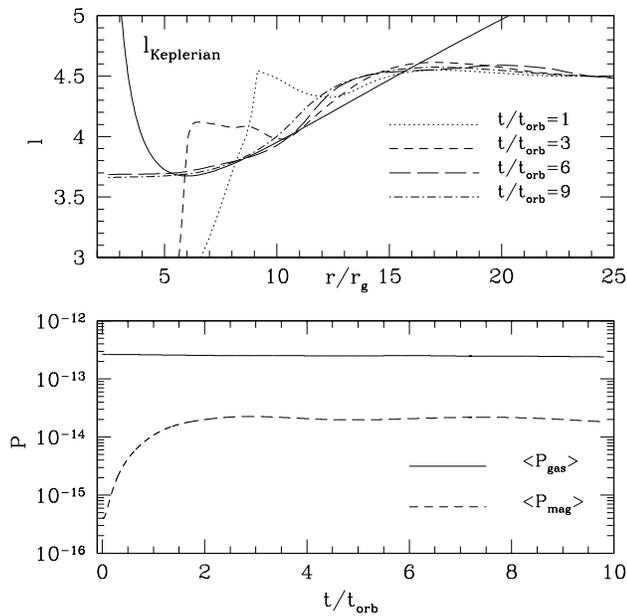,angle=0,width=8.5cm}}
\caption{
\label{fig9}
Top panel: the equatorial angular momentum approaches
 the Keplerian profile as an effect of the
magnetorotational instability. Bottom panel: time
evolution of the total gas pressure (solid line) and of
 the magnetic pressure (dashed line). 
}
\end{figure}

The four panels of Fig.~\ref{fig8} display isocontours of
the rest mass density,  logarithmically spaced, during
the first few orbital periods of the evolution. These
results correspond to a
simulation employing the HLL solver with a computational
grid of 200 radial  zones and 100 angular zones. It was
first shown by \cite{balbus} that the dynamics  of such
magnetized thick disks is governed by the 
so-called  magnetorotational instability (MRI), which
generates turbulence in the disk and  helps explaining
the transport of angular momentum outwards. In
axisymmetry the development of
the MRI is much less significant than in full  three
dimensions and 
manifests itself through the appearence of the so-called
``channel  solution'' \citep{devilliers:03}. 
This feature of the solution becomes visible in our simulation  after about
three orbital periods, as shown in Fig.~\ref{fig8}, in the
form of a  high density elongated structure near the
equatorial plane.
We report in Fig.~\ref{fig9}
two additional distinctive features that can be
unambiguosly attributed to the MRI. The first one,
showed in the top panel, represents the
transport of angular  momentum (initially constant)
outward, which acquires a Keplerian profile
(indicated by a thick solid line) as the
evolution proceeds. Correspondingly, the botton panel
shows the rapid increase of the
(mean) magnetic pressure
(dashed line) with respect to the gas pressure (solid line)
during the first two orbital periods and due to the MRI
driven turbulence.

%We are developping a parallel version of the code
%in order to 
%to evolve efficiently 
%additional simulations with higher resolutions and
%with increasingly larger values of the magnetization
%parametere. With that, 

We note, however, that the present status of the 
numerical code
does not allow us to evolve efficiently 
additional simulations with higher resolutions and
with increasingly larger values of the magnetization
parameter. As a result, 
the typical  distorsion of the
isodensity contours   produced by the MRI
is not visible in Fig.~\ref{fig8}.
A parallel version of the code is currently
under development. This will allow for higher resolution
simulations of magnetized disks in  astrophysical
contexts. 

%%%%%%%%%%%%%%%%%%%%%%%%%%%%%%%%%%%%%%%%%%%%%%%%%%
\section{Conclusions}
%%%%%%%%%%%%%%%%%%%%%%%%%%%%%%%%%%%%%%%%%%%%%%%%%%

In this paper we have presented a procedure to solve numerically the general 
relativistic magnetohydrodynamic equations within the framework of the $3+1$ 
formalism.  The work reported here represents the extension of our previous 
investigation \citep{banyuls:97} where magnetic fields
were not considered. The  GRMHD equations have been
explicitely written in conservation form to exploit
their hyperbolic character in the solution procedure
using Riemann solvers. Most of the theoretical ingredients which
are necessary in order to build up high-resolution
shock-capturing schemes based on the solution of local
Riemann problems have been  discussed. 
%In particular, the
%characteristic information of the Jacobian matrices  of
%the system of equations, i.e. the eigenvalues and
%right-eigenvectors, has been  discussed and implemented
%in a numerical code in three alternative HRSC schemes,
%either upwind as HLL or Roe, or symmetric as KT. 
In particular, we have described and implemented  
three alternative HRSC schemes,
either upwind as HLL and Roe, or symmetric as KT. 
Our implementation of the Roe-type  Riemann solver has made
use of the equivalence principle of general relativity
which allows to use, locally, the characteristic
information of the system of  equations in the special
relativistic limit, following a slight modification of
the procedure first presented in \citet{pons:98}. Further
information regarding  the renormalization of the
eigenvectors of the GRMHD flux-vector Jacobians has  been
deferred to an accompanying paper \citep{anton05}. The
work reported in this  paper, hence, follows the recent
stir of activity in the ongoing efforts of developing
robust numerical codes for the GRMHD system of equations,
as exemplified by the  investigations presented in the
last few years by a number of groups  \citep{devilliers1,gammie:03,duez05,komissarov05}.

Our formulation of the equations and numerical procedure have been assessed by 
performing the various test simulations discussed in earlier works in the 
literature, including magnetized shock tubes in flat
spacetimes, spherical accretion  onto a Schwarzshild
black hole, equatorial magnetized accretion in the Kerr
spacetime,  as well as evolution of thick accretion disks
subject to the development of the  magnetorotational
instability. The code has proved to be second order
accurate and  has successfully passed all considered
tests. In the near future we plan to apply  this code in
a number of astrophysical scenarios involving compact
objects where both strong gravitational fields and
magnetic fields need be taken into account. 

\section*{Acknowledgments}
This research has been supported by the Spanish
Ministerio de Educaci\'on y Ciencia (grant
AYA2004-08067-C03-01,  AYA2004-08067-C03-02 and
SB2002-0128).  The computations were
performed on the Beowulf Cluster  for Numerical
Relativity {\it ``Albert100''} at the University of Parma
and on the  SGI/Altix3000 computer {\it ``CERCA''} at the
Servicio de Inform\'atica de la Universidad de Valencia.

%%%%%%%%%%%%%%%%%%%%%%%%%%%%%%%%%%%%%
\appendix
\section{Magnetized Michel accretion}
\label{app_A}
%%%%%%%%%%%%%%%%%%%%%%%%%%%%%%%%%%%%%

In this Appendix we prove that there is not a consistent solution 
for a force-free magnetic field added to the spherically symmetric accretion 
of a perfect fluid onto a Schwarzschild black hole. In general, it is not at 
all obvious that a hydrodynamical solution can be ``dressed'' with a force-free 
magnetic field. \cite{oron:02} has shown that the form of the four-current 
compatible with a force-free magnetic field is given by
\begin{equation}
\label{current}
{\cal J}^\mu=\rho_q u^\mu + \eta b^\mu
\end{equation}
where $\rho_q$ is the proper charge density. 
Note that when $\eta=0$, i.e.~when 
the current is only due to the convective term, the
assumption of force-free is  automatically guaranteed by
the ideal MHD condition. However, we will consider  here
the more general expression given by
Eq.~(\ref{current}). If we write  explicitely the four
vanishing components of the electric field in the
comoving  frame of the accreting fluid, $F_{\mu\nu}u^\nu=0$,
recalling that the velocity field is  given by
$u^\mu=(u^0,u^1,u^2,u^3)=(u^t,u^r,0,0)$, we find 
\begin{eqnarray}
\label{FF1}
F_{01}&=&0, \\
\label{FF2}
F_{02}u^0+F_{12}u^1&=&0, \\
\label{FF3}
F_{31}&=&0,
\end{eqnarray}
where we have also used the fact that $F_{03}=\partial_0 A_3 - \partial_3 A_0=0$. 
Let us next consider the first couple of Maxwell equations
\begin{equation}
\label{max1}
F_{[\alpha\beta,\gamma]}=0,
\end{equation}
where the comma denotes partial differentiation. After writing them explicitly for 
all possible combinations we obtain
\begin{eqnarray}
F_{01,2} + F_{12,0} + F_{20,1} &=& 0, \\
F_{01,3} + F_{13,0} + F_{30,1} &=& 0, \\
F_{02,3} + F_{23,0} + F_{30,2} &=& 0, \\
F_{12,3} + F_{23,1} + F_{31,2} &=& 0 \ .
\end{eqnarray}
By the symmetries of the spacetime and by relations (\ref{FF1})-(\ref{FF3}) this 
system reduces to  
\begin{eqnarray}
\label{thetadependence1}
F_{02,1}&=&0, \\
\label{thetadependence2}
F_{23,1}&=&0. 
\end{eqnarray}
Summarizing, among the 6 components of the antisymmetric
electromagnetic tensor  $F_{\mu\nu}$, 3 of them
vanish, namely $F_{01}=F_{03}=F_{13}=0$. Among the
remaining 3, only two are independent, since the
constraint (\ref{FF2}) has to be fulfilled. Furthermore,
according to Eqs.~(\ref{thetadependence1}) and
(\ref{thetadependence2}), $F_{02}$ and $F_{23}$ are
functions of the angle  $\theta$ only,
$F_{02}=F_{02}(\theta)$ and $F_{23}=F_{23}(\theta)$, and
are  therefore constants along fluid lines. Taking all
this into account we can write  the components of the
magnetic field explicitly, using definition (\ref{bmu})
in the main text
\begin{eqnarray}
\label{BB0}
b^0 &=&\frac{1}{\sqrt{-g}} F_{23}u_1,\\
\label{BB1}
b^1 &=&-\frac{1}{\sqrt{-g}} F_{23}u_0=\frac{b^0 u_0}{u_1} \\
\label{BB2}
b^2 &=& 0, \\
\label{BB3}
b^3 &=&\frac{1}{\sqrt{-g}} ( F_{02}u_1 - F_{12}u_0)=-\frac{F_{02}}{\sqrt{-g}u^1}.
\end{eqnarray}
Note that Eq.~(\ref{BB3}) can be alternatively computed from the condition 
$b^\mu u_\mu=0$. 

Up to this point we have shown that the magnetic
field is completely determined by two constants, $F_{23}$
and $F_{02}$. We now consider the second couple of
Maxwell equations, namely $\nabla_\nu
F^{\mu\nu}=4\pi {\cal J}^\mu$. 
According to the assumption on the four-current,
Eq.~(\ref{current}), and on the  four-velocity in the
case of spherical accretion, these equations become 
\begin{eqnarray}
\label{mm0}
\partial_2(\sqrt{-g}F^{02})&=&4\pi\sqrt{-g}(\rho_q u^0 + \eta b^0), \\
\label{mm1}
\partial_2(\sqrt{-g}F^{12})&=&4\pi\sqrt{-g}(\rho_q u^1 + \eta b^1), \\ 
\label{mm2}
\partial_1(\sqrt{-g}F^{21})&=& 0, \\
\label{mm3}
\partial_2(\sqrt{-g}F^{32})&=&4\pi\sqrt{-g}\eta b^3 \ , 
\end{eqnarray}
where $F^{02}=F_{02}/g_{00} g_{22}$ and
$F^{12}=F_{12}/g_{11} g_{22}$. 
From (\ref{mm2})  it follows that the term
$F_{12}/g_{11}$ must be a function of the angular
coordinate  $\theta$ only which, recalling (\ref{FF2})
and the fact that both $u^0$ and $u^1$ are  functions of
$r$, implies that $F_{12}=F_{02}=0$. As a result, the
toroidal component of the magnetic field $b^3$
vanishes. Moreover, according to Eq.~(\ref{mm3}), the
term  $F_{23}/(r^2 \sin\theta)$ must be a function of $r$
only. Given that $F_{23}= F_{23}(\theta)$, it must be
$F_{23}= A \sin\theta$, with $A$ a constant. Finally,
(\ref{mm0}) and (\ref{mm1}) are now reduced to the
following homogeneous system in  the unknowns $\rho_q$
and $\eta$ 
\begin{eqnarray}
u^0 \rho_q  + b^0 \eta  = 0, \\
u^1 \rho_q  + b^1 \eta  = 0. 
\end{eqnarray}
Imposing the vanishing of the determinant gives $b^0/b^1=u^0/u^1$, which cannot be 
satisfied since it violates the constraint coming from the combination of the 
orthogonality condition $b^\mu u_\mu=0$ and the normalization condition 
$u^\mu u_\mu=-1$. This concludes the proof that it is not possible to add a force-free 
magnetic field to the hydrodynamic solution of spherical accretion in a Schwarzschild 
spacetime that satisfies the full set of Maxwelll equations.

%%%%%%%%%%%%%%%%%%%%%%%%%%%%%%%%%%%%%%%%%
\appendix
\section{Magnetized thick accretion disk}
\label{app_B}
%%%%%%%%%%%%%%%%%%%%%%%%%%%%%%%%%%%%%%%%%

In this Appendix we show that it is not possible to build
a consistent stationary  and axisymmetric solution for a
magnetized torus by simply adding a force-free magnetic
field to the hydrodynamic equilibrium model of an
isoentropic thick accretion disk  \citep{kow:78,font:02a}. The proof, that for simplicity we limit to the case of 
Schwarzschild spacetime but can be extended to a Kerr black hole as well, could 
follow the same reasoning of the previous Appendix. However, the demonstration is 
more direct if one exploits some topological properties of the expected solution. 
In fact, from Maxwell equations it is possible to show that the magnetic field of a 
perfectly conducting medium endowed with a purely toroidal motion has to be purely
poloidal, i.e.~$ b^r \neq 0$, $b^\theta \neq 0$, while
$b^t=b^\phi=0$. 
Under these conditions 
the magnetic field lines lie on the surfaces of constant magnetic potential $A_\phi$ 
(magnetic surfaces), which coincide with the surfaces of constant angular velocity 
$\Omega=u^\phi/u^t$. This property prevents the generation of a toroidal component 
of the magnetic field, even in the presence of differential rotation (Ferraro's 
theorem), and allows to introduce a new coordinate system $(x_1,x_2)$ such that 
$x_1$ varies along the poloidal field lines and $x_2$ is constant along them 
\citep{oron:02}. In this new coordinate system the magnetic field will only have 
one non-vanishing component $b^{1}$, while $b^{2}=0$.

According to \citet{bekenstein}, for a force-free magnetic field in an isoentropic
flow the quantity $U=u^t u_\phi$ is constant along the magnetic surfaces, and it can 
be used to define the new coordinate $x_2$. In the case of circular motion in 
Schwarzschild spacetime this quantity reads 
\begin{equation}
U=-\frac{\Omega g_{\phi\phi}}{g_{tt}(1-\Omega \ell)}
\end{equation}
where $\ell=-u_\phi/u_t$ is the specific angular momentum. According to von Zeipel's 
theorem \citep{vonzeipel} $\ell$ is constant along surfaces of constant $\Omega$ for the 
class of barotropic hydrodynamic models that we are considering. Therefore, both 
$\Omega$ and $\ell$ are constant along magnetic surfaces and the new coordinate $x_2$ 
can be defined as
\begin{eqnarray}
\label{x2}
x_2 = \left(\frac{U}{\Omega}(1-\Omega \ell)\right)^{1/2} =
\left(-\frac{g_{\phi\phi}}{g_{tt}}\right)^{1/2}=\frac{r
\sin\theta}{(1-\frac{2M}{r})^{1/2}} , \ 
\end{eqnarray}
which is the so-called von Zeipel parameter \citep{chak:85}. The other coordinate $x_1$ 
can be chosen such that orthogonality between $x_1$ and $x_2$ is preserved, i.e. 
$g_{12}=0$. After some calculations involving straightforward metric coefficient 
transformations, this choice yields to
\begin{eqnarray}
\label{x1}
x_1 = (r-3M)\cos\theta \ .
\end{eqnarray}
In computing Eq.~(\ref{x1}) we have made the reasonable ansatz that $x_1$ is factorized 
as $x_1=p(r)q(\theta)$. \cite{oron:02} has shown that, in order to satisfy the second 
couple of Maxwell's equations and the scalar equation $\nabla_\mu(h b^\mu)=0$, which 
can be proved to hold for any isoentropic magnetized flow, the following factorization 
in terms of generic functions of $x_1$ and $x_2$ must exist
\begin{equation}
\label{cond}
\frac{g_{11}}{g_{22}\Delta (u^t)^4} = f(x_1) h(x_2) ,
\end{equation}
where $\Delta=-g_{tt}g_{\phi\phi}$ in the Schwarzschild
metric. 
From the normalization
condition $u^\mu u_\mu=-1$ it follows that
$(u^t)^2=1/ [(g_{tt}(1-x_2^2\Omega^2)]$, and
Eq.~(\ref{cond}) becomes 
\begin{equation}
\label{qui}
\left(1-\frac{2M}{r}\right)^2 (x_2)^2 (1-\Omega^2 (x_2)^2)^{-2} = f(x_1) h(x_2) .
\end{equation}
Since $\Omega=\Omega(x_2)$, 
Eq.~(\ref{qui}) requires that
the term $1-2M/r$ is
factorizable as $f(x_1)  h(x_2)$, which can be shown not
to be possible. Hence, the constraint (\ref{cond}) cannot
be met, and a force-free magnetized torus built from the
isoentropic hydrodynamic model  of a thick accretion disk
cannot be obtained.  

\bibliographystyle{apj}
\bibliography{ms}

\begin{thebibliography}{67}
\expandafter\ifx\csname natexlab\endcsname\relax\def\natexlab#1{#1}\fi

\bibitem[{Anile(1989)}]{anile}
Anile, A.~M. 1989, Relativistic fluids and magneto-fluids (Cambridge, England:
  Cambridge University Press)

\bibitem[{Ant\'on {et~al.}(2005)Ant\'on, Mart\'{\i}, Miralles, \&
  Ib\'a\~nez}]{anton05}
Ant\'on, L., Mart\'{\i}, J.~M., Miralles, J.~A., \& Ib\'a\~nez, J.~M. 2005, in
  preparation

\bibitem[{Arnowitt {et~al.}(1962)Arnowitt, Deser, \& Misner}]{ADM}
Arnowitt, R., Deser, S., \& Misner, C.~W. The Dynamics of General Relativity,
  ed. L.~Witten (New York: John Wiley), 227--265

\bibitem[{{Balbus} \& {Hawley}(1991)}]{balbus}
{Balbus}, S.~A. \& {Hawley}, J.~F. 1991, \apj, 376, 214

\bibitem[{Balsara(2001)}]{balsara01}
Balsara, D. 2001, \apjs, 132, 83

\bibitem[{{Banyuls} {et~al.}(1997){Banyuls}, {Font}, {Ib\'a\~nez},
  {Mart\'{\i}}, \& {Miralles}}]{banyuls:97}
{Banyuls}, F., {Font}, J.~A., {Ib\'a\~nez}, J.~M., {Mart\'{\i}}, J.~M., \&
  {Miralles}, J.~A. 1997, \apj, 476, 221

\bibitem[{{Baumgarte} \& {Shapiro}(2003)}]{baumgarte1}
{Baumgarte}, T.~W. \& {Shapiro}, S.~L. 2003, \apj, 585, 921

\bibitem[{Bekenstein \& Oron(1979)}]{bekenstein}
Bekenstein, J.~D. \& Oron, E. 1979, \prd, 19, 2827

\bibitem[{{Blandford} \& {Payne}(1982)}]{blandford:82}
{Blandford}, R.~D. \& {Payne}, D.~G. 1982, \mnras, 199, 883

\bibitem[{{Blandford} \& {Znajek}(1977)}]{blandford:77}
{Blandford}, R.~D. \& {Znajek}, R.~L. 1977, MNRAS, 179, 433

\bibitem[{{Bocquet} {et~al.}(1995){Bocquet}, {Bonazzola}, {Gourgoulhon}, \&
  {Novak}}]{bocquet}
{Bocquet}, M., {Bonazzola}, S., {Gourgoulhon}, E., \& {Novak}, J. 1995, \aap,
  301, 757

\bibitem[{{Brio} \& {Wu}(1988)}]{brio}
{Brio}, M. \& {Wu}, C.~C. 1988, J.~Comput.~Phys., 75, 400

\bibitem[{{Chakrabarti}(1985)}]{chak:85}
{Chakrabarti}, S.~K. 1985, \apj, 288, 1

\bibitem[{{De Villiers} \& {Hawley}(2003{\natexlab{a}})}]{devilliers1}
{De Villiers}, J. \& {Hawley}, J.~F. 2003{\natexlab{a}}, \apj, 589, 458

\bibitem[{{De Villiers} \& {Hawley}(2003{\natexlab{b}})}]{devilliers:03}
---. 2003{\natexlab{b}}, \apj, 592, 1060

\bibitem[{{Del Zanna} {et~al.}(2003){Del Zanna}, {Bucciantini}, \&
  {Londrillo}}]{delzanna}
{Del Zanna}, L., {Bucciantini}, N., \& {Londrillo}, P. 2003, A\&A, 400, 397

\bibitem[{{Duez} {et~al.}(2005){Duez}, {Liu}, {Shapiro}, \&
  {Stephens}}]{duez05}
{Duez}, M.~D., {Liu}, Y.~T., {Shapiro}, S.~L., \& {Stephens}, B.~C. 2005,
  astro-ph/0503420

\bibitem[{Evans \& Hawley(1988)}]{evans88}
Evans, C. \& Hawley, J.~F. 1988, \apj, 332, 659

\bibitem[{{Fishbone} \& {Moncrief}(1976)}]{fish:76}
{Fishbone}, L.~G. \& {Moncrief}, V. 1976, \apj, 207, 962

\bibitem[{{Font}(2003)}]{fontlr}
{Font}, J.~A. 2003, Living Reviews in Relativity, 6, 4

\bibitem[{{Font} \& {Daigne}(2002)}]{font:02a}
{Font}, J.~A. \& {Daigne}, F. 2002, MNRAS, 334, 383

\bibitem[{{Font} {et~al.}(1994){Font}, {Ibanez}, {Marquina}, \&
  {Marti}}]{font:94}
{Font}, J.~A., {Ibanez}, J.~M., {Marquina}, A., \& {Marti}, J.~M. 1994, \aap,
  282, 304

\bibitem[{{Font} {et~al.}(2000){Font}, {Miller}, {Suen}, \& {Tobias}}]{font:00}
{Font}, J.~A., {Miller}, M., {Suen}, W., \& {Tobias}, M. 2000, \prd, 61, 044011

\bibitem[{{Gammie}(1999)}]{gammie}
{Gammie}, C.~F. 1999, \apj, 522, L57

\bibitem[{{Gammie} {et~al.}(2003){Gammie}, {McKinney}, \& {T{\'
  o}th}}]{gammie:03}
{Gammie}, C.~F., {McKinney}, J.~C., \& {T{\' o}th}, G. 2003, \apj, 589, 444

\bibitem[{Harten {et~al.}(1983)Harten, Lax, \& van Leer}]{harten:83}
Harten, A., Lax, P.~D., \& van Leer, B. 1983, SIAM Review, 25, 35

\bibitem[{{Ibanez} {et~al.}(2001){Ibanez}, {Aloy}, {Font}, {Marti}, {Miralles},
  \& {Pons}}]{ibanez:01}
{Ibanez}, J.~M., {Aloy}, M.~A., {Font}, J.~A., {Marti}, J.~M., {Miralles},
  J.~A., \& {Pons}, J.~A. 2001, in Godunov Methods: Theory and Applications,
  ed. E.F.Toro (New York: Kluver Academic), 485--496

\bibitem[{{Koide}(2003)}]{koide03}
{Koide}, S. 2003, \prd, 67, 104010

\bibitem[{Koide {et~al.}(2000)Koide, Meier, Shibata, \& Kudoh}]{koide00}
Koide, S., Meier, D.~L., Shibata, K., \& Kudoh, T. 2000, \apj, 536, 668

\bibitem[{Koide {et~al.}(1998)Koide, Shibata, \& Kudoh}]{koide98}
Koide, S., Shibata, K., \& Kudoh, T. 1998, \apj, 495, L63

\bibitem[{Koide {et~al.}(2002)Koide, Shibata, Kudoh, \& Meier}]{koide02a}
Koide, S., Shibata, K., Kudoh, T., \& Meier, D.~L. 2002, Science, 295, 1688

\bibitem[{{Koldoba} {et~al.}(2002){Koldoba}, {Kuznetsov}, \&
  {Ustyugova}}]{koldoba}
{Koldoba}, A.~V., {Kuznetsov}, O.~A., \& {Ustyugova}, G.~V. 2002, MNRAS, 333,
  932

\bibitem[{Komissarov(1999)}]{komissarov99}
Komissarov, S.~S. 1999, MNRAS, 303, 343

\bibitem[{{Komissarov}(2005)}]{komissarov05}
{Komissarov}, S.~S. 2005, \mnras, 359, 801

\bibitem[{{Kouveliotou} {et~al.}(1998){Kouveliotou}, {Dieters}, {Strohmayer},
  {van Paradijs}, {Fishman}, {Meegan}, {Hurley}, {Kommers}, {Smith}, {Frail},
  \& {Murakami}}]{kouveliotou}
{Kouveliotou}, C., {Dieters}, S., {Strohmayer}, T., {van Paradijs}, J.,
  {Fishman}, G.~J., {Meegan}, C.~A., {Hurley}, K., {Kommers}, J., {Smith}, I.,
  {Frail}, D., \& {Murakami}, T. 1998, \nat, 393, 235

\bibitem[{{Kozlowski} {et~al.}(1978){Kozlowski}, {Jaroszynski}, \&
  {Abramowicz}}]{kow:78}
{Kozlowski}, M., {Jaroszynski}, M., \& {Abramowicz}, M.~A. 1978, \aap, 63, 209

\bibitem[{Kurganov \& Tadmor(2000)}]{tadmor}
Kurganov, A. \& Tadmor, E. 2000, J.~Comput.\ Phys., 160, 214

\bibitem[{Leismann {et~al.}(2005)Leismann, Ant\'on, Aloy, M\"uller, Mart\'{\i},
  Miralles, \& Ib\'a\~nez}]{tobias}
Leismann, T., Ant\'on, L., Aloy, M.~A., M\"uller, E., Mart\'{\i}, J.~M.,
  Miralles, J.~A., \& Ib\'a\~nez, J.~M. 2005, A\&A, in press

\bibitem[{LeVeque(1998)}]{leveque}
LeVeque, R.~J. 1998, in Computational methods for astrophysical fluid flow.
  Saas-Fee Advanced Course 27, ed. O.~Steiner \& A.~Gautschy (Berlin, Germany:
  Springer), 1--159

\bibitem[{{Londrillo} \& {del Zanna}(2004)}]{londrillo:04}
{Londrillo}, P. \& {del Zanna}, L. 2004, J.~Comput.~Phys., 195, 17

\bibitem[{Lucas-Serrano {et~al.}(2004)Lucas-Serrano, Font, Ib\'a\~nez, \&
  Mart\'{\i}}]{arturo}
Lucas-Serrano, A., Font, J.~A., Ib\'a\~nez, J.~M., \& Mart\'{\i}, J.~M. 2004,
  A\&A, 428, 703

\bibitem[{{Mart{\'{\i}}} \& {M{\" u}ller}(2003)}]{martilr:03}
{Mart{\'{\i}}}, J.~M. \& {M{\" u}ller}, E. 2003, Living Reviews in Relativity,
  6, 7

\bibitem[{{McKinney} \& {Gammie}(2004)}]{mckinney:04}
{McKinney}, J.~C. \& {Gammie}, C.~F. 2004, \apj, 611, 977

\bibitem[{Michel(1972)}]{michel:72}
Michel, F. 1972, Astrophys. Spa. Sci., 15, 153

\bibitem[{{Oron}(2002)}]{oron:02}
{Oron}, A. 2002, \prd, 66, 023006

\bibitem[{Papadopoulos \& Font(1999)}]{papadopoulos}
Papadopoulos, P. \& Font, J.~A. 1999, \prd, 61, 024015

\bibitem[{Penrose(1969)}]{penrose}
Penrose, R. 1969, Nuovo Cimento, 1, 252

\bibitem[{{Pons} {et~al.}(1998){Pons}, {Font}, {Ib\'a\~nez}, {Mart\'{\i}}, \&
  {Miralles}}]{pons:98}
{Pons}, J.~A., {Font}, J.~A., {Ib\'a\~nez}, J.~M., {Mart\'{\i}}, J.~M., \&
  {Miralles}, J.~A. 1998, A\&A, 339, 638

\bibitem[{Romero {et~al.}(2005)Romero, Mart\'{\i}, Pons, Miralles, \&
  Ib\'a\~nez}]{romero}
Romero, R., Mart\'{\i}, J.~M., Pons, J.~A., Miralles, J.~A., \& Ib\'a\~nez,
  J.~M. 2005, J.~Fluid Mech., accepted

\bibitem[{{Ryu} {et~al.}(1995){Ryu}, {Jones}, \& {Frank}}]{ryu:95}
{Ryu}, D., {Jones}, T.~W., \& {Frank}, A. 1995, \apj, 452, 785

\bibitem[{{Ryu} {et~al.}(1998){Ryu}, {Miniati}, {Jones}, \& {Frank}}]{ryu:98}
{Ryu}, D., {Miniati}, F., {Jones}, T.~W., \& {Frank}, A. 1998, \apj, 509, 244

\bibitem[{Shibata \& Font(2005)}]{shibata}
Shibata, M. \& Font, J.~A. 2005, \prd, submitted

\bibitem[{Shibata \& Sekiguchi(2005)}]{shibata:05}
Shibata, M. \& Sekiguchi, Y. 2005, \prd, submitted

\bibitem[{{Shu} \& {Osher}(1988)}]{shu:88}
{Shu}, C. \& {Osher}, S. 1988, J.~Comput.~Phys., 77, 439

\bibitem[{Sloan \& Smarr(1985)}]{sloan85}
Sloan, J. \& Smarr, L.~L. General relativistic magnetohydrodynamics, ed. J.~L.
  J.~Centrella \& R.~Bowers (Boston: Jones and Bartlett), 52--68

\bibitem[{{T{\' o}th}(2000)}]{toth:00}
{T{\' o}th}, G. 2000, J.~Comput.~Phys., 161, 605

\bibitem[{Takahashi {et~al.}(1990)Takahashi, Nitta, Tatematsu, \&
  Tomimatsu}]{takahashi}
Takahashi, M., Nitta, S., Tatematsu, Y., \& Tomimatsu, A. 1990, \apj, 363, 206

\bibitem[{Toro(1997)}]{toro:97}
Toro, E.~F. 1997, Riemann solvers and numerical methods for fluid dynamics
  (Berlin: Springer Verlag)

\bibitem[{van Putten(1991)}]{vanputten91}
van Putten, M.~H.~P.~M. 1991, Comm.~Math.~Phys., 141, 63

\bibitem[{van Putten(1993)}]{vanputten93}
---. 1993, J.~Comput.~Phys., 99, 341

\bibitem[{von Zeipel(1924)}]{vonzeipel}
von Zeipel, H. 1924, \mnras, 84, 665

\bibitem[{Wilson(1979)}]{wilson79}
Wilson, J.~R. A numerical method for relativistic hydrodynamics, ed. L.~Smarr
  (Cambridge: Cambridge University Press), 423--445

\bibitem[{Yokosawa(1993)}]{yokosawa93}
Yokosawa, M. 1993, PASJ, 45, 207

\bibitem[{{Yokosawa}(1995)}]{yokosawa95}
{Yokosawa}, M. 1995, PASJ, 47, 605

\bibitem[{Yokosawa \& Inui(2005)}]{yokosawa05}
Yokosawa, M. \& Inui, T. 2005, astro-ph/0504024

\bibitem[{{Zanotti} {et~al.}(2003){Zanotti}, {Rezzolla}, \&
  {Font}}]{zanotti:03}
{Zanotti}, O., {Rezzolla}, L., \& {Font}, J.~A. 2003, \mnras, 341, 832

\bibitem[{{Zhang}(1989)}]{zhang}
{Zhang}, X. 1989, \prd, 39, 2933

\end{thebibliography}

\end{document}